\documentclass[a4paper,12pt]{article}
\pdfoutput=1
\usepackage{graphicx}
\usepackage{mathtools}
\usepackage{amssymb,mathrsfs,amsmath}
\usepackage{amsfonts}
\usepackage[footnotesize]{caption}
\usepackage[font=scriptsize]{subcaption}
\usepackage{color}
\usepackage{xcolor}
\usepackage{braket}
\usepackage{cite}
\usepackage{bbm}
\usepackage{url}
\usepackage[colorlinks, linkcolor=red, anchorcolor=black, citecolor=green]{hyperref}
\usepackage{multirow}
\usepackage{relsize}
\usepackage{fullpage}
\usepackage{makecell}
\usepackage{blkarray}
\usepackage{fullpage}
\usepackage{float}
\restylefloat{table}

\usepackage[T1]{fontenc}

\usepackage{makecell}

\def\simge{\mathrel{
   \rlap{\raise 0.511ex \hbox{$>$}}{\lower 0.511ex \hbox{$\sim$}}}}

\def\simle{\mathrel{
   \rlap{\raise 0.511ex \hbox{$<$}}{\lower 0.511ex \hbox{$\sim$}}}}

\def\s#1{\setbox0=\hbox{$#1$}
\rlap{\ifdim\wd0>.7em\kern.22\wd0\else\kern.1\wd0\fi /}#1}

\newcommand{\newc}{\newcommand}
\newc{\be}{\begin{equation}}
\newc{\ee}{\end{equation}}
\newc{\bea}{\begin{eqnarray}}
\newc{\eea}{\end{eqnarray}}
\newc{\ben}{\begin{equation*}}
\newc{\een}{\end{equation*}}
\newc{\bean}{\begin{eqnarray*}}
\newc{\eean}{\end{eqnarray*}}
\newc{\ol}{\overline}
\newc{\wt}{\widetilde}
\newc{\bs}{\boldsymbol}
\newc{\m}{\mathcal}
\newc{\la}{\lambda}
\newc{\lra}{\longrightarrow}
\newc{\vp}{\varphi}
\newc{\ti}{\tilde}

\newc{\simlt}{~\mbox{\smaller\(\lesssim\)}~}
\newc{\simgt}{~\mbox{\smaller\(\gtrsim\)}~}

\begin{document}

\begin{titlepage}
\begin{flushright}
\hfill\mbox{{\small\tt USTC-ICTS/PCFT-21-13}} \\[5mm]
\begin{minipage}{0.2\linewidth}
\normalsize
\end{minipage}
\end{flushright}
\begin{center}
{\bf\Large
\boldmath{Modular $S_4\times SU(5)$ GUT }
} \\[12mm]
Gui-Jun~Ding$^{\dagger,\ddagger}$\footnote{E-mail: {\tt
dinggj@ustc.edu.cn}},  \
Stephen~F.~King$^{\star}$
\footnote{E-mail: \texttt{king@soton.ac.uk}; ORCID: https://orcid.org/0000-0002-4351-7507},\
Chang-Yuan~Yao$^{\S}$\footnote{E-mail: {\tt
yaocy@nankai.edu.cn}},
\\[-2mm]
\end{center}
\vspace*{0.50cm}

\centerline{$^{\dagger}$ \it
Peng Huanwu Center for Fundamental Theory, Hefei, Anhui 230026, China}
\vspace*{2mm}
\centerline{$^{\ddagger}$ \it
Interdisciplinary Center for Theoretical Study and Department of Modern Physics,}
\centerline{\it
University of Science and Technology of China, Hefei, Anhui 230026, China}
\vspace*{2mm}
\centerline{$^{\star}$ \it
Department of Physics and Astronomy, University of Southampton,}
\centerline{\it
SO17 1BJ Southampton, United Kingdom }
\vspace*{2mm}
\centerline{$^{\S}$ \it
School of Physics, Nankai University, Tianjin 300071, China}
\vspace*{1.20cm}

\begin{abstract}
{\noindent
Modular symmetry offers the possibility to provide an origin of discrete flavour symmetry and to break it along particular symmetry preserving directions without introducing flavons or driving fields. It is also possible to use a weighton field to account for charged fermion mass hierarchies rather than a Froggatt-Nielsen mechanism. Such an approach can be applied to flavoured Grand Unified Theories (GUTs) which can be greatly simplified using modular forms. As an example, we consider a modular version of a previously proposed $S_4\times SU(5)$ GUT, with Gatto-Sartori-Tonin and Georgi-Jarlskog relations, in which all flavons and driving fields are removed, with their effect replaced by modular forms with moduli assumed to be at various fixed points, rendering the theory much simpler. In the neutrino sector there are two right-handed neutrinos constituting a Littlest Seesaw model satisfying Constrained Sequential Dominance (CSD) where the two columns of the Dirac neutrino mass matrix are proportional to $(0,1, -1)$ and $(1, n, 2-n)$ respectively, and $n=1+\sqrt{6}\approx 3.45$ is prescribed by the modular symmetry, with predictions subject to charged lepton mixing corrections. We perform a numerical analysis, showing quark and lepton mass and mixing correlations around the best fit points.
}
\end{abstract}
\end{titlepage}

\section{Introduction}

The mystery of the three families of quarks and leptons, and their patterns of masses and mixings, including in particular the origin of tiny neutrino mass with large mixing, remains a good motivation for studying physics beyond the Standard Model (BSM). The quest for unification, in order to understand the quantum numbers of quarks and leptons within a family, including charge quantisation and anomaly cancellation, and the desire to unify the three gauge forces, also motivates BSM studies.

The combination of family symmetry and grand unified theories (GUTs)~\cite{Georgi:1974sy} provides a powerful and constrained framework~\cite{King:2017guk}. The minimal $SU(5)$ GUT symmetry \cite{Georgi:1974sy} allows neutrino mass and mixing to be included by the addition of any number of right-handed neutrinos, where the type I seesaw mechanism~\cite{Minkowski:1977sc, Yanagida:1979as, GellMann:1980vs, Glashow:1979nm, Mohapatra:1979ia, Schechter:1980gr, Schechter:1981cv} explains the smallness of neutrino masses compared to charged lepton masses. For example, $SU(5)$ can be combined with the minimal type I seesaw mechanism which includes just two right-handed neutrinos (2RHN)~\cite{King:1999mb,Frampton:2002qc}.

A class of highly predictive 2RHN models are based on constrained sequential dominance (CSD)~\cite{King:2005bj,Antusch:2011ic,King:2013iva,King:2015dvf,King:2016yvg,Ballett:2016yod,King:2018fqh,King:2013xba,King:2013hoa,Bjorkeroth:2014vha}. The CSD scheme (also called Littlest Seesaw \cite{King:2015dvf}) assumes that the two columns of the Dirac neutrino mass matrix are proportional to $(0,1, -1)$ and $(1, n, 2-n)$ respectively in the RHN diagonal basis, where $n$ may take any value, giving rise to large class of CSD($n$) models.  In all such models, the lepton mixing matrix is predicted to be the trimaximal TM$_1$ pattern, in which the first column of the lepton mixing matrix follows the tri-bimaximal mixing pattern while allowing for a non-zero $U_{e3}$, and the neutrino masses are normal ordered with the lightest neutrino being massless with $m_1=0$.  The phenomenologically successful cases include CSD($3$)~\cite{King:2013iva,King:2015dvf,King:2016yvg,Ballett:2016yod,King:2018fqh} and CSD($4$)~\cite{King:2013xba,King:2013hoa}. For example CSD($3$) has been shown to originate from $S_4$~\cite{King:2016yvg}.

It has been suggested that flavour symmetry groups such as $S_4$ could originate from the quotient group of the modular group $SL(2,\mathbb{Z})$ over the principal congruence subgroups~\cite{Feruglio:2017spp}, where the leptons are assigned to have modular weights, the Yukawa and mass parameters are modular forms, namely holomorphic functions of a complex modulus $\tau$, with even (or odd) modular weights. The modular forms of level $N$ and integer weight $k$ can be arranged into some modular multiplets of the inhomogeneous finite modular group $\Gamma_N\equiv\overline{\Gamma}/\overline{\Gamma}(N)$ if $k$ is an even number~\cite{Feruglio:2017spp}. For example, $\Gamma_2\cong S_3$~\cite{Kobayashi:2018vbk,Kobayashi:2018wkl,Kobayashi:2019rzp,Okada:2019xqk}, $\Gamma_3\cong A_4$~\cite{Feruglio:2017spp,Criado:2018thu,Kobayashi:2018vbk,Kobayashi:2018scp,deAnda:2018ecu,Okada:2018yrn,Kobayashi:2018wkl,Novichkov:2018yse,Nomura:2019jxj,Okada:2019uoy,Nomura:2019yft,Ding:2019zxk,Okada:2019mjf,Nomura:2019lnr,Kobayashi:2019xvz,Asaka:2019vev,Gui-JunDing:2019wap,Zhang:2019ngf,Nomura:2019xsb,Wang:2019xbo,Kobayashi:2019gtp,King:2020qaj,Ding:2020yen,Okada:2020rjb,Nomura:2020opk,Asaka:2020tmo,Okada:2020brs,Yao:2020qyy,Feruglio:2021dte}, $\Gamma_4\cong S_4$~\cite{Penedo:2018nmg,Novichkov:2018ovf,deMedeirosVarzielas:2019cyj,Kobayashi:2019mna,King:2019vhv,Criado:2019tzk,Wang:2019ovr,Gui-JunDing:2019wap,Wang:2020dbp}, $\Gamma_5\cong A_5$~\cite{Novichkov:2018nkm,Ding:2019xna,Criado:2019tzk} and $\Gamma_7\cong PSL(2, Z_7)$~\cite{Ding:2020msi} cases have been studied.  We shall be interested in the $\Gamma_4\cong S_4$ modular symmetry in the present work. If the modular weight $k$ is an odd positive number, the modular forms can be decomposed into multiplets of the homogeneous finite modular group $\Gamma'_N\equiv\Gamma/\Gamma(N)$ which is the double cover of $\Gamma_N$~\cite{Liu:2019khw}. Notice that top-down constructions in string theory usually give rise to $\Gamma'_N$~\cite{Baur:2019kwi,Baur:2019iai}.  All modular forms of integral weight are polynomials of weight one modular forms, the odd weight and even weight modular forms are in the representations $\rho_{\mathbf{r}}(S^2)=-1$ and $\rho_{\mathbf{r}}(S^2)=+1$ respectively. The modular invariant approach based on $\Gamma'_3\cong T'$~\cite{Liu:2019khw,Lu:2019vgm}, $\Gamma'_4\cong S'_4$~\cite{Novichkov:2020eep,Liu:2020akv} and $\Gamma'_5\cong A'_5$~\cite{Wang:2020lxk,Yao:2020zml} has been studied to understand the flavor structure of quarks and leptons. Furthermore, the case that the modular weight $k$ is a rational number has been explored~\cite{Liu:2020msy,Yao:2020zml}, then $(c\tau+d)^k$ is not the automorphy factor and certain multiplier is necessary. As a consequence, the modular symmetry and finite modular groups should be extended to their metaplectic covers~\cite{Liu:2020msy,Yao:2020zml}. The predictive power of the modular invariance can be considerably improved by including the generalized CP symmetry~\cite{Novichkov:2019sqv,Baur:2019kwi,Baur:2019iai}.  The framework of modular invariant supersymmetric theory has been extended to incorporate several moduli for both factorizable~\cite{deMedeirosVarzielas:2019cyj} and non-factorizable~\cite{Ding:2020zxw,Ding:2021iqp} cases.

The complex modulus $\tau$ is restricted to complex values in the upper half complex plane, but it can take special values, associated with residual symmetries of the finite modular group, where such values are called stabilisers~\cite{Novichkov:2018yse,Novichkov:2018ovf,Gui-JunDing:2019wap,Novichkov:2021evw}. In the framework of string theory, there may be enhanced symmetries at various points in moduli space, which allows for various different stabilisers occurring simultaneously within the low energy effective theory \cite{Nilles:2020nnc}. This is known as ``Local Flavour Unification''.  Another ingredient in realistic models is to include quark mass and mixing, which has been addressed in non-GUT models in \cite{Okada:2019uoy,Lu:2019vgm,King:2020qaj,Liu:2020akv,Yao:2020zml}. It is remarkable that modular invariance can also address the origin of mass hierarchies without introducing an additional Froggatt-Nielsen (FN) $U(1)$~\cite{Froggatt:1978nt} symmetry.  The role of the FN flavon is played by a singlet field called the weighton~\cite{King:2020qaj}, which carries a non-zero modular weight, but no other charges. The question of quark and lepton mass and mixing in the context of modular $SU(5)$ GUTs was first studied in an $(\Gamma_3\simeq A_4)\times SU(5)$ model in~\cite{deAnda:2018ecu}, then $(\Gamma_2\simeq S_3)\times SU(5)$~\cite{Kobayashi:2019rzp,Du:2020ylx}, and $(\Gamma_4\simeq S_4)\times SU(5)$~\cite{Zhao:2021jxg,King:2021fhl}. Most recently a comprehensive analysis has been performed on $(\Gamma_3\simeq A_4)\times SU(5)$ models \cite{Chen:2021zty}, without restricting the modulus $\tau$ to take any special values.

In this paper, we shall propose a new modular model based on $(\Gamma_4\simeq S_4)\times SU(5)$ which exploits the large range of stabilisers studied in \cite{Gui-JunDing:2019wap}. In particular, we shall show that the minimal 2RHN seesaw model based on CSD($n$) with $n=1+\sqrt{6}\approx 3.45$, intermediate between CSD($3$) and CSD($4$), can be incorporated into an $SU(5)$ GUT, where it is however subject to charged lepton corrections. We shall also include a weighton field \cite{King:2020qaj} to ameliorate the large hierarchies in the charged fermion mass matrices, although some tuning will remain at the per cent level. Using the stabilisers, we are able to reproduce some of the classic features of GUT models such as the Gatto-Sartori-Tonin (GST)~\cite{Gatto:1968ss} and Georgi-Jarlskog (GJ)~\cite{Georgi:1979df} relations, although we shall see that these relations apply in a more generalised form as the limiting cases of a choice of parameters.  Indeed in the case of GJ, this is to be welcomed, since those relations do not work if strictly imposed.

It is interesting to compare the present model with the $S_4\times SU(5)\times U(1)$ GUT model of Hagedorn, King and Luhn (HKL) \cite{Hagedorn:2010th,Hagedorn:2012ut} (see also \cite{Dimou:2015yng,Dimou:2015cmw}) where the GST and GJ relations are used.  Many features of the HKL model may be reproduced without any flavon fields, by replacing the flavons by modular forms of $S_4$ with several moduli assumed to be at their fixed points, thereby drastically simplifying the model.  In particular the HKL model requires 9 flavons as compared to the single weighton $\phi$ in the modular model. The HKL model also requires 13 driving fields, to drive and align the flavon VEVs, while the modular model only requires one driving field $\chi$, illustrating the dramatic simplification of family GUT models in the framework of modular symmetry.

The layout of the remainder of the paper is as follows.  In section~\ref{sec:modularform} we review modular invariance, while in section~\ref{level4} we focus on modular forms of $\Gamma_4\cong S_{4}$ at level 4. In section~\ref{sec:model2} we present the model based on modular $S_{4} \times SU(5)$ GUT, including the fields and the fixed points that we assume, as well as the resulting Yukawa matrices and the neutrino mass matrices satisfying CSD($3.45$). In section~\ref{numerical} we discuss the numerical results of the model, based on a weighton expansion parameter of $0.1$, which ameliorates the hierarchies in the quark and charged lepton masses, leaving a residual fine tuning at the per cent level, and perform a scan about a best fit point, focussing on the charged lepton corrections to CSD($3.45$) neutrino mixing. Section~\ref{conclusion} concludes the main body of the paper. In Appendix~\ref{sec:S4_group_app} we summarise some relevant aspects of the group theory of $\Gamma_4\cong S_{4}$. The results of fit at two local minima of $\chi^2$ are given in Appendix~\ref{sec:app-local-minima}.

\section{\label{sec:modularform} The modular invariance approach }
In this section we recapitulate the concept of modular symmetry and the formalism of modular invariant supersymmetric theories~\cite{Ferrara:1989bc,Ferrara:1989qb,Feruglio:2017spp}. The modular group $SL(2, \mathbb{Z})$ often denoted as $\Gamma$ is the group of $2\times2$ matrices with integer coefficients and unit determinant under matrix multiplication~\cite{diamond2005first},
\begin{equation}
SL(2, \mathbb{Z})=\left\{
\begin{pmatrix}
a  &  b  \\
c  &  d
\end{pmatrix}\Big|a,b,c,d\in\mathbb{Z}, ad-bc=1 \right\}\,.
\end{equation}
The modular group $SL(2, \mathbb{Z})$ is an infinite discrete group, generated by two generators $S$ and $T$ with
\begin{equation}
S=\left(\begin{array}{cc}
0 & 1 \\
-1  & 0
\end{array}
\right),~~~\quad T=\left(\begin{array}{cc}
1 & 1 \\
0  & 1
\end{array}
\right)\,,
\end{equation}
which satisfy the multiplication rules
\begin{equation}
\label{eq:ST-relations}S^4=(ST)^3=1,~~~S^2T=TS^2\,.
\end{equation}
Under the modular group $\Gamma$, the complex modulus $\tau$ in the upper half plane with $\texttt{Im}(\tau)>0$ transforms as
\begin{equation}
\tau\rightarrow\gamma\tau=\frac{a\tau+b}{c\tau+d}=\gamma(\tau),~~~\gamma=\begin{pmatrix}
a  &  b  \\
c  &  d
\end{pmatrix}\in\Gamma\,.
\end{equation}
It is easy to show
\begin{equation}
\gamma\gamma'(\tau)=\gamma(\gamma'(\tau)),~~~~~\texttt{Im}(\gamma\tau)=\frac{\texttt{Im}(\tau)}{|c\tau+d|^2}\,.
\end{equation}
Since $\pm\gamma$ induce the same transformation on $\tau$, we get a transformation group $SL(2,\mathbb{Z})/\left\{1, S^2\right\}\equiv \overline{\Gamma}$ which is called inhomogeneous modular group. The pair of matrices $\gamma$ and $-\gamma$ are considered to be identical. For $N=1$ global supersymmetry, the most general form of the action is
\begin{equation}
\mathcal{S}=\int d^4x d^2\theta d^2\bar{\theta}\, \mathcal{K}(\Phi, \bar{\Phi}, \tau, \bar{\tau})+\left[\mathcal{W}(\Phi, \tau)+\text{H.c.}\right]\,,
\end{equation}
where $\mathcal{K}$ is the K\"ahler potential, $\mathcal{W}$ is the superpotential, and $\Phi$ collectively denotes chiral superfields of the theory and they are separated into sectors $\varphi^{(I)}$. Under the action of modular group $\Gamma$, the supermultiplets $\varphi^{(I)}$ of each sector are assumed to transform as following
\begin{equation}
\varphi^{(I)}\rightarrow (c\tau+d)^{-k_{I}}\rho_{I}(\gamma)\varphi^{(I)}\,,
\end{equation}
where $-k_{I}$ is the modular weight, and $\rho_{I}(\gamma)$ is the unitary representation of the quotient group $\Gamma_N=\overline{\Gamma}/\overline{\Gamma}(N)$~\cite{Feruglio:2017spp} or its double cover $\Gamma'_N=\Gamma/\Gamma(N)$~\cite{Liu:2019khw}. Here $\Gamma(N)$ is the principal congruence subgroup of level $N$,
\begin{equation}
\Gamma(N)=\left\{\begin{pmatrix}
a  & b\\
c   & d
\end{pmatrix}\Big|a=d=1(\texttt{mod}~N), b=c=0\,(\texttt{mod}~N)\right\}\,.
\end{equation}
The groups $\overline{\Gamma}(N)$ are slightly different from $\Gamma(N)$ with $\overline{\Gamma}(N)=\Gamma(N)/\left\{1, S^2\right\}$ for $N=1, 2$ and $\overline{\Gamma}(N)=\Gamma(N)$ for $N>2$. Notice that $T^N\in\Gamma(N)$ and consequently the homogeneous finite modular group $\Gamma'_N$ can be expressed in terms of the modular generators $S$ and $T$ satisfying the relations in Eq.~\eqref{eq:ST-relations} together with $T^N=1$, while the multiplication rules of $\Gamma_N$ are $S^2=(ST^3)=T^N$ for $N\leq5$. The K\"ahler potential $\mathcal{K}$  is a real gauge-invariant function of the chiral superfields $\Phi$, the modulus field $\tau$ and their conjugates. Following~\cite{Feruglio:2017spp}, we choose a minimal form of the K\"ahler potential,
\begin{equation}
\label{eq:kahler}\mathcal{K}(\Phi,\bar{\Phi},\tau,\bar{\tau}) = -h\Lambda^2\log(-i\tau+i\bar{\tau})+\sum_I(-i\tau+i\bar{\tau})^{-k_I}|\varphi^{(I)}|^2\,,
\end{equation}
where $h$ is a positive constant and $\Lambda$ is the cutoff scale. After the modulus $\tau$ gets a vacuum expectation, this K\"ahler potential gives the kinetic terms for the scalar components of the supermultiplet $\Phi_I$ and the modulus field $\tau$ as follows
\begin{equation}
\frac{h\Lambda^2}{\langle-i\tau+i\bar{\tau}\rangle^2}\partial_{\mu}\bar{\tau}\partial^{\mu}\tau+\sum_I\frac{\partial_{\mu}\bar{\varphi}^{(I)}\partial^{\mu}\varphi^{(I)}}{\langle-i\tau+i\bar{\tau}\rangle^{k_I}}
\end{equation}
Notice that the K\"ahler potential can not be fixed by the modular symmetry~\cite{Chen:2019ewa}, for instance the operators $(-i\tau+i\overline{\tau})^{k-k_I}\left(\varphi^{(I)\dagger}Y^{(k)\dagger}_{\mathbf{r}}Y^{(k)}_{\mathbf{r}}\varphi^{(I)}\right)_{\mathbf{1}}$ for any integer $k$ and irreducible representation $\mathbf{r}$ are always compatible with modular symmetry. The additional terms with additional parameters in the K\"ahler potential can reduce the predictive power of the modular invariance approach~\cite{Chen:2019ewa}. In top-down construction, the modular symmetry always appears with traditional flavor symmetry, and the modular weights and K\"ahler potential as well as the superpotential would be strongly constrained. In particularly the K\"ahler potential is found to coincide with the minimal one in Eq.~\eqref{eq:kahler} after rescaling of fields is considered~\cite{Baur:2019kwi,Baur:2019iai,Nilles:2020nnc,Nilles:2020kgo}.
The superpotential $\mathcal{W}$ is a holomorphic gauge-invariant function of the chiral superfields $\varphi^{(I)}$ and $\tau$. It can be expanded into power series of supermultiplets $\Phi_I$
\begin{equation}
\mathcal{W}(\Phi,\tau)=\sum_n Y_{I_1...I_n}(\tau)\varphi^{(I_1)}...\varphi^{(I_n)}\,.
\end{equation}
Each term of the superpotential should be invariant under the modular transformation, thus $Y_{I_1...I_n}(\tau)$ should be modular forms of weight $k_Y$ and level $N$ and it transforms in the representation $\rho_Y$ of the finite modular group $\Gamma_N$ or $\Gamma'_N$:
\begin{equation}
Y(\tau)\to Y(\gamma\tau)=(c\tau+d)^{k_Y}\rho_{Y}(\gamma)Y(\tau)\,.
\end{equation}
If the finite modular group is $\Gamma_N$, $k_Y$ must be an even integer while odd $k_Y$ is allowed for the double cover groups $\Gamma'_N$~\cite{Liu:2019khw}. The modular forms of level $N$ span a linear space of finite dimension for each nonnegative weight $k$. The integral (even) weight modular forms can be generated from the tensor products of modular forms of weight one (two). Modular invariance of the superpotential requires the modular weights and representations should fulfill the conditions:
\begin{equation}
k_Y=k_{I_1}+...+k_{I_n},~\quad~ \rho_Y\otimes\rho_{I_1}\otimes\ldots\otimes\rho_{I_n}\ni\mathbf{1}\,.
\end{equation}

\section{\label{level4} Modular forms of $\Gamma_4\cong S_{4}$ at level 4 }

In this section, we review the construction of the modular forms of level 4 from the products of $\eta(\tau)$~\cite{Gui-JunDing:2019wap}, where $\eta(\tau)$ is the Dedekind eta function,
\begin{equation}
\eta(\tau)=q^{1/24} \prod_{n=1}^\infty \left(1-q^n \right),~~~~ q=e^{2\pi i\tau}\,.
\end{equation}
It is easy to compute numerically by writing the eta function in power series,
\begin{equation}
\eta(\tau)=q^{1/24}\sum^{+\infty}_{n=-\infty} (-1)^nq^{n(3n-1)/2}\,.
\end{equation}
The linear space of modular forms of weight $k$ and level 4 has dimension $2k+1$. In the present work, we are concerned with the inhomogeneous finite modular group $\Gamma_4\cong S_4$, the weights of the modular forms have to be even. There are five independent weight 2 modular forms at level 4 and they can be arranged into a doublet $\mathbf{2}$ and a triplet $\mathbf{3}$ of the finite modular group $S_4$~\cite{Gui-JunDing:2019wap},
\begin{equation}
Y^{(2)}_{\mathbf{2}}(\tau)=\begin{pmatrix}
Y_1(\tau) \\
Y_2(\tau)
\end{pmatrix}\,,
\quad
Y^{(2)}_{\mathbf{3}}(\tau)=\begin{pmatrix}
Y_3(\tau) \\
Y_4(\tau) \\
Y_5(\tau)
\end{pmatrix}\,,
\end{equation}
where
\begin{eqnarray}
\nonumber
Y_1(\tau)&=&16\omega^2e_1(\tau)-8(2+\omega^2)e_3(\tau)+\omega^2e_5(\tau),\\
\nonumber
Y_2(\tau)&=&16e_1(\tau)+8i\sqrt{3}\,e_3(\tau)+e_5(\tau), \\
\nonumber
Y_3(\tau)&=&-\omega^2\left[16e_1(\tau)+16(1-i)e_2(\tau)+4(1+i)e_4(\tau) -e_5(\tau)\right], \\
\nonumber
Y_4(\tau)&=&-\omega\left[16e_1(\tau)+8(1-\sqrt{3})(-1+i)e_2(\tau)-2(1+\sqrt{3})(1+i)e_4(\tau)-e_5(\tau)\right],\\
Y_5(\tau)&=&-16e_1(\tau)+8(1+\sqrt{3})(1-i)e_2(\tau)+2(1-\sqrt{3})(1+i)e_4(\tau)+e_5(\tau)\,,
\end{eqnarray}
with $\omega=e^{2\pi i/3}$ and
\begin{equation}
e_i(\tau)=\frac{\eta^{12-4i}(4\tau)\eta^{6i-10}(2\tau)}{\eta^{2i-2}(\tau)}\,.
\end{equation}
The modular forms of weight 2 and level 4 can also be constructed from the linear combinations of $\frac{\eta'(\tau/4)}{\eta(\tau/4)}$, $\frac{\eta'((\tau+1)/4)}{\eta((\tau+1)/4)}$, $\frac{\eta'((\tau+2)/4)}{\eta((\tau+2)/4)}$, $\frac{\eta'((\tau+3)/4)}{\eta((\tau+3)/4)}$, $\frac{\eta'(4\tau)}{\eta(4\tau)}$ and $\frac{\eta'(\tau+1/2)}{\eta(\tau+1/2)}$~\cite{Penedo:2018nmg,Novichkov:2018ovf}, where $\eta'(\tau)$ denotes the derivative of $\eta(\tau)$ with respect to $\tau$. The two construction methods give the same $q-$expansion of modular forms up to an overall factor. Note that the two triplets $\mathbf{3}$ and $\mathbf{3}'$ in our basis correspond to $\mathbf{3}'$ and $\mathbf{3}$ of~\cite{Penedo:2018nmg,Novichkov:2018ovf} respectively. The higher weight modular forms can be expressed as polynomials of $Y_{1,2,3,4,5}(\tau)$. The linearly independent weight 4 modular forms can be arranged into a singlet $\mathbf{1}$, a doublet $\mathbf{2}$ and two triplets $\mathbf{3}$, $\mathbf{3'}$ of $S_4$ as follows
\begin{eqnarray}
\nonumber&& Y^{(4)}_{\mathbf{1}}=\left(Y^{(2)}_{\mathbf{2}}Y^{(2)}_{\mathbf{2}}\right)_{\mathbf{1}}= 2Y_1Y_2 , \\
\nonumber&& Y^{(4)}_{\mathbf{2}}=\left(Y^{(2)}_{\mathbf{2}}Y^{(2)}_{\mathbf{2}}\right)_{\mathbf{2}}=
\begin{pmatrix}
Y^2_2 \\
Y^2_1
\end{pmatrix}, \\
\nonumber&& Y^{(4)}_{\mathbf{3}}=\left(Y^{(2)}_{\mathbf{2}}Y^{(2)}_{\mathbf{3}}\right)_{\mathbf{3}}=\begin{pmatrix}
Y_1Y_4+Y_2Y_5 \\
Y_1Y_5+Y_2Y_3  \\
Y_1Y_3+Y_2Y_4
\end{pmatrix}, \\
&& Y^{(4)}_{\mathbf{3'}}=\left(Y^{(2)}_{\mathbf{2}}Y^{(2)}_{\mathbf{3}}\right)_{\mathbf{3'}}=\begin{pmatrix}
Y_1Y_4-Y_2Y_5 \\
Y_1Y_5-Y_2Y_3 \\
Y_1Y_3-Y_2Y_4
\end{pmatrix} \,.
\end{eqnarray}
There are 13 linearly independent weight 6 modular forms which decompose as $\mathbf{1}\oplus\mathbf{1'}\oplus\mathbf{2}\oplus\mathbf{3}\oplus\mathbf{3}\oplus\mathbf{3'}$ under $S_4$:
\begin{eqnarray}
\nonumber&&Y^{(6)}_{\mathbf{1}}=\left(Y^{(2)}_{\mathbf{2}}Y^{(4)}_{\mathbf{2}}\right)_{\mathbf{1}}= Y^3_1+Y^3_2 , \\
\nonumber&&Y^{(6)}_{\mathbf{1'}}=\left(Y^{(2)}_{\mathbf{2}}Y^{(4)}_{\mathbf{2}}\right)_{\mathbf{1'}}= Y^3_1-Y^3_2, \\
\nonumber&&Y^{(6)}_{\mathbf{2}}=\left(Y^{(2)}_{\mathbf{2}}Y^{(4)}_{\mathbf{1}}\right)_{\mathbf{2}}=
\begin{pmatrix}
2Y^2_1Y_2\\
2Y_1Y^2_2
\end{pmatrix}\,, \\
\nonumber&&Y^{(6)}_{\mathbf{3}, I}=\left(Y^{(2)}_{\mathbf{3}}Y^{(4)}_{\mathbf{1}}\right)_{\mathbf{3}}=
\begin{pmatrix}
2 Y_1Y_2Y_3 \\
2Y_1Y_2Y_4 \\
2Y_1Y_2Y_5
\end{pmatrix}\,, \\
\nonumber&&Y^{(6)}_{\mathbf{3}, II}=\left(Y^{(2)}_{\mathbf{3}}Y^{(4)}_{\mathbf{2}}\right)_{\mathbf{3}}=\begin{pmatrix}
Y_1^2Y_5+Y_2^2Y_4 \\
Y_1^2Y_3+Y_2^2Y_5 \\
Y_1^2 Y_4 +Y_2^2Y_3
\end{pmatrix}\,,\\
&&Y^{(6)}_{\mathbf{3'}}=\left(Y^{(2)}_{\mathbf{3}}Y^{(4)}_{\mathbf{2}}\right)_{\mathbf{3'}}=
-\begin{pmatrix}
Y^2_1Y_5-Y^2_2Y_4 \\
Y^2_1Y_3-Y^2_2Y_5 \\
Y^2_1Y_4-Y^2_2Y_3
\end{pmatrix}\,.
\end{eqnarray}
Higher weight modular forms can be built in the same fashion, see Refs.~\cite{Penedo:2018nmg,Novichkov:2018ovf} for modular forms of weight 8 and weight 10. Note that in our working basis the representation matrices are different from those of~\cite{Penedo:2018nmg,Novichkov:2018ovf}, and they are related to the choices of~\cite{Penedo:2018nmg,Novichkov:2018ovf} by unitary transformations. The allowed and forbidden representations for modular forms of a given (even) weight are summarised in table~\ref{tab:allowed}, not only for $k_Y=2,4,6$ above, but also for $k_Y=8,10$~\cite{Novichkov:2018ovf}.

\begin{table}[t!]
\renewcommand{\tabcolsep}{0.5mm}
\begin{center}
\begin{tabular}{|c|c|c|}\hline\hline
Weight $k_Y$ & Allowed $\mathbf{r}$ & Forbidden $\mathbf{r}$   \\ \hline

$0$  & $\mathbf{1}$ &  $\mathbf{1'}$,$\mathbf{2}$,$\mathbf{3}$,$\mathbf{3'}$
\\ \hline

$2$  & $\mathbf{2}$,$\mathbf{3}$ &  $\mathbf{1}$,$\mathbf{1'}$,$\mathbf{3'}$
\\ \hline

$4$  &  $\mathbf{1}$,$\mathbf{2}$,$\mathbf{3}$,$\mathbf{3'}$  & $\mathbf{1'}$
\\ \hline

$6$  &  $\mathbf{1}$,$\mathbf{1'}$,$\mathbf{2}$,$\mathbf{3}$,$\mathbf{3'}$  &  ---
\\ \hline

$8$  &  $\mathbf{1}$,$\mathbf{2}$,$\mathbf{3}$,$\mathbf{3'}$  &  $\mathbf{1'}$
\\ \hline

$10$  &  $\mathbf{1}$,$\mathbf{1'}$,$\mathbf{2}$,$\mathbf{3}$,$\mathbf{3'}$  &  ---
\\ \hline\hline

\end{tabular}
\caption{\label{tab:allowed} Summary of the allowed and forbidden modular forms $Y^{(k_Y)}_{\mathbf{r}}$ for a given (even) weight $k_Y=0,2,4,6,8,10$.  The allowed and forbidden $S_4$ representations $\mathbf{r}=\mathbf{1},\mathbf{1'},\mathbf{2},\mathbf{3},\mathbf{3'}$ are shown in each case. Explicit modular forms are given in the main text for weights $k_Y=2,4,6$. }
\end{center}
\end{table}

If a modulus parameter $\tau_0$ is invariant under the action of a nontrivial $SL(2, \mathbb{Z})$ transformation $\gamma_0\neq\pm I$, we call $\tau_0$ the fixed point of $\gamma_0$, where $\gamma_0$ is the stabilizer of $\tau_0$, i.e.
\begin{equation}
\label{eq:fp_equa}\gamma_0\tau_0=\tau_0\,.
\end{equation}
We show the alignments of the modular forms for three nontrivial fixed points in the fundamental domain in table~\ref{tab:residual-fixed-points},
\begin{equation}
\tau_S=i,~~~\tau_{ST}=\omega= -\frac{1}{2}+i\frac{\sqrt{3}}{2},~~~\tau_{TS}=-\omega^2= \frac{1}{2}+i\frac{\sqrt{3}}{2}.
\end{equation}

\begin{table}[hpb!]
\centering
\begin{tabular}{|c|c|c|c|} \hline  \hline
& $\tau_{S}=i$ & $\tau_{ST}=\omega$ & $\tau_{TS}=-\omega^2$  \\ \hline
$Y^{(2)}_{\mathbf{2}}$ & $Y_{S}(1,-1)$ & $Y_{ST}(0,1)$ & $Y_{TS}(1,0)$ \\ \hline
$Y^{(2)}_{\mathbf{3}}$ & $-\frac{\omega}{\sqrt{3}}Y_{S}(1, 1+\sqrt{6}, 1-\sqrt{6}) $ & $ \sqrt{3}\omega Y_{ST}(0, 1, 0)$ & $-\frac{2\omega}{\sqrt{3}}Y_{TS}(1,1,-\frac{1}{2}) $  \\
\hline \hline
$Y^{(4)}_{\mathbf{1}}$ & $-2Y_{S}^{2}$ & $0$ & $0$ \\ \hline
$Y^{(4)}_{\mathbf{2}}$ & $Y_{S}^{2}(1,1)$ & $Y_{ST}^{2}(1,0)$ & $Y_{TS}^{2}(0,1)$ \\ \hline
$Y^{(4)}_{\mathbf{3}}$ & $-2\sqrt{2}\omega Y_{S}^{2} (1,-\frac{1}{2},-\frac{1}{2})$ & $\sqrt{3}\omega Y_{ST}^{2}(0, 0, 1) $ & $-\frac{2\omega}{\sqrt{3}} Y_{TS}^{2} (1, -\frac{1}{2}, 1)$  \\
\hline
$Y^{(4)}_{\mathbf{3'}}$ & $-\frac{2\omega}{\sqrt{3}} Y_{S}^{2} (1, 1-\sqrt{\frac{3}{2}}, 1+\sqrt{\frac{3}{2}})$ & $-\sqrt{3}\omega  Y_{ST}^{2} (0,0,1)$ & $-\frac{2\omega}{\sqrt{3}} Y_{TS}^{2} (1, -\frac{1}{2}, 1) $  \\ \hline \hline
$Y^{(6)}_{\mathbf{1}}$ & $0$ & $Y_{ST}^{3}$ & $Y_{TS}^{3}$ \\ \hline
$Y^{(6)}_{\mathbf{1'}}$ & $2Y_{S}^{3}$ & $-Y_{ST}^{3}$ & $Y_{TS}^{3}$ \\ \hline
$Y^{(6)}_{\mathbf{2}}$ & $-2Y_{S}^{3}(1,-1)$ & $(0,0)$ & $(0,0)$ \\ \hline
$Y^{(6)}_{\mathbf{3}, I}$ & $\frac{2\omega}{\sqrt{3}} Y_{S}^{3} (1, 1+\sqrt{6}, 1-\sqrt{6})$ & $(0,0,0)$ & $(0,0,0)$ \\ \hline
$Y^{(6)}_{\mathbf{3}, II}$ & $-\frac{2\omega}{\sqrt{3}} Y_{S}^{3} (1, 1-\sqrt{\frac{3}{2}}, 1+\sqrt{\frac{3}{2}})$ & $\sqrt{3}\omega Y_{ST}^{3} (1,0,0)$ & $\frac{\omega}{\sqrt{3}}Y_{TS}^{3}(1,-2,-2)$  \\ \hline
$Y^{(6)}_{\mathbf{3'}}$ & $-2 \sqrt{2}\omega Y_{S}^{3} (1,-\frac{1}{2}, -\frac{1}{2})$ & $\sqrt{3}\omega Y_{ST}^{3} (1,0,0)$ & $-\frac{\omega}{\sqrt{3}}Y_{TS}^{3}(1,-2,-2)$  \\ \hline \hline
\end{tabular}
\caption{\label{tab:residual-fixed-points}The values of the modular forms with weights $k_Y=2,4,6$ and level 4 at the fixed points $\tau_{S}$, $\tau_{ST}$ and $\tau_{TS}$, where $Y_{S}\simeq-1.045-0.603i$, $Y_{ST}\simeq1.793$ and $Y_{TS}\simeq-0.896-1.553i$. Notice that there are two linearly independent modular forms in the representations $\mathbf{3}$ and $\mathbf{3'}$ at weight 8~\cite{Penedo:2018nmg,Novichkov:2018ovf}, and  both $Y^{(8)}_{\mathbf{3}}(\tau_{ST})$ and $Y^{(8)}_{\mathbf{3'}}(\tau_{ST})$ are proportional to $(0, 1, 0)^{T}$. Analogously there are three independent triplet modular forms $Y^{(10)}_{\mathbf{3}}$ aligned in the direction $(0, 0, 1)$ at the point $\tau=\tau_{ST}$.}
\end{table}

In addition, we shall be interested in using the following fixed point outside the fundamental domain, at $\gamma \tau_{S}=2+i$
~\cite{Gui-JunDing:2019wap},
\begin{equation}
Y^{(4)}_\mathbf{3}(2+i)\propto (0,1,-1),\ \ Y^{(6)}_\mathbf{3'}(2+i)\propto (0,1,-1).
\label{2+i}
\end{equation}
There are many more possible fixed points~\cite{Gui-JunDing:2019wap}, but these shown in Eq.~\eqref{2+i} and table~\ref{tab:residual-fixed-points} are the only ones that we shall need for the model construction, to which we turn in the next section.

\section{\label{sec:model2} Modular $S_{4} \times SU(5)$ GUT }

In this section, we shall construct a $SU(5)$ GUT model based on $S_4$ modular symmetry, and no flavon field other than the modulus $\tau$ is used.

\subsection{Fields and Symmetries}

The model is based on the grand unified group $SU(5)$ combined with modular $S_4$ family symmetry, with the field content shown in table~\ref{tab:field-content-4th}. The left-handed quarks and leptons are unified into the representations $\mathbf{\overline{5}}$,
$\mathbf{10}$ and $\mathbf{1}$ of $SU(5)$ according to
\begin{align}
F_{\alpha} \sim \overline{\bf 5} \sim  \left( \begin{array}{c} d_r^c \\ d_b^c \\ d_g^c \\ e^- \\ -\nu \end{array} \right)_{\!\!\alpha} \,,\qquad
T_{\alpha} \sim {\bf 10} \sim  \left( \begin{array}{ccccc} 0 & u_g^c & -u_b^c & u_r & d_r \\ . & 0 & u_r^c & u_b & d_b \\
 . & . & 0 & u_g & d_g \\ . & . & . & 0 & e^c \\ . & . & . & . & 0 \end{array}\right)_{\!\!\alpha} \,, \qquad
N_{\rm a, s} \sim {\bf 1}\,,
\label{Eqn:SU5_reps}
\end{align}
where the fields with superscript $c$ stands for CP conjugated fields (which would be right-handed without the $c$ operation), and $\alpha=1,\ldots , 3$ is the family index. The three families are controlled by a family symmetry $S_4$, with $F$ forming a triplet and the first two families of $T$ forming a doublet, while the third family $T_3$ (containing the top quark) is a singlet, as summarized in table~\ref{tab:field-content-4th}. The choice of the third family $T_3$ being a singlet, permits a renormalisable top quark Yukawa coupling to the singlet Higgs discussed below. There are two (CP conjugated) right-handed neutrinos which transform under $S_4$ as $N_{\rm a, s}\sim \mathbf{1'},\mathbf{1}$.

The $S_4$ singlet Higgs fields $H_5, \ H_{\overline{5}}$ and $H_{\overline{45}}$, each contain a doublet $SU(2)_L \times U(1)_Y$ representation that eventually form the standard up ($H_u$) and down ($H_d$) Higgses of the Minimal Supersymmetric Standard Model (MSSM), where the $H_d$ emerges as a linear combination of doublets from the $H_{\overline{5}}$ and $H_{\overline{45}}$~\cite{Chung:2003fi}\footnote{As $H_{\bf{\overline{5}}}$ and $H_{\bf{\overline{45}}}$ transform differently under $U(1)$, it is clear that the mechanism which spawns the low energy Higgs doublet $H_d$ must necessarily break $U(1)$. Although the discussion of any details of the $SU(5)$ GUT symmetry breaking (which, e.g., could even have an extra dimensional origin) are beyond the scope of our paper, we remark that a mixing of $H_{\bf{\overline{5}}}$ and   $H_{\bf{\overline{45}}}$ could be induced by introducing the pair $H^\pm_{\bf{24}}$ with $U(1)$ charges $\pm 1$ in addition to the standard $SU(5)$ breaking Higgs $H^0_{\bf{24}}$.}. The VEVs of the two neutral Higgses are:
\begin{eqnarray}
\upsilon_u&=&\frac{\upsilon}{\sqrt{1+t_\beta^2}}t_\beta,~~~~~~\upsilon_d=\frac{\upsilon}{\sqrt{1+t_\beta^2}},
\end{eqnarray}\\[-3mm]
where $t_\beta\equiv \tan\beta=\frac{\upsilon_u}{\upsilon_d}$ and $\upsilon = \sqrt{\upsilon_u^2+\upsilon_d^2}=174$ GeV.

\begin{table}[hptb]
\begin{center}
\begin{tabular}{|c||c|c|c|c|c||c|c|c||c|c|}
\hline\hline
Field & $T_3$ & $T=(T_2, T_1)^{T}$ & $F$ & $N_{\rm a}$ & $N_{\rm s}$ & $H_5$ & $H_{\overline{5}}$ & $H_{\overline{45}}$ & $\phi$ & $\chi^0$
\\ \hline
$SU(5)$ & $\mathbf{10}$ & $\mathbf{10}$ & $\mathbf{\overline{5}}$ & $\mathbf{1}$ & $\mathbf{1}$ & $\mathbf{5}$ & $\mathbf{\overline{5}}$ & $\mathbf{\overline{45}}$ & $\mathbf{1}$  & $\mathbf{1}$
\\      \hline
$S_4$ & $\mathbf{1}$ & $\mathbf{2}$ & $\mathbf{3}$ & $\mathbf{1}$ & $\mathbf{1'}$ & $\mathbf{1'}$ & $\mathbf{1}$ & $\mathbf{1}$ & $\mathbf{1}$ & $\mathbf{1}$
\\ \hline
$k_I$ & $4$ & $1$ & $3$ & $4$ & $-1$ & $-2$ & $1$ & $1$ & $1$  & $0$
\\ \hline\hline
\end{tabular}
\caption{\label{tab:field-content-4th} The modular $S_4\times SU(5)$ model with matter and Higgs fields and their associated representations and modular weights given by $-k_I$. We have also included a weighton field $\phi$, a driving field $\chi^0$. }
\end{center}
\end{table}

\subsection{The weighton}
The weighton was introduced in~\cite{King:2020qaj} as a means of naturally generating fermion mass hierarchies. The weighton will develop a vacuum expectation value (VEV) which may be driven by a leading order superpotential term
\begin{equation}
\mathcal{W}_{driv}= \chi^0(Y^{(4)}_{\mathbf{1}}\frac{\phi^4}{M_{fl}^2} -M^2)\,,
\label{eq:driving}
\end{equation}
where $\chi^0$ is an $S_4$ singlet driving superfield with zero modular weight, while $M$ is a free dimensionful mass scale, where we assume $M\ll M_{fl}$. This is similar to the usual driving field mechanism familiar from flavon models~\cite{Altarelli:2010gt,Ishimori:2010au,King:2013eh,King:2014nza,King:2015aea,King:2017guk,Feruglio:2019ktm}, except for the presence of the lowest weight singlet modular form $Y^{(4)}_{\mathbf{1}}$ listed in Eq.~\eqref{eq:driving}, where the quadratic term $\phi^2$ is forbidden since $Y^{(2)}_{\mathbf{1}}$ does not exist, and we have dropped higher powers such as $\phi^{6}$, and so on. As usual~\cite{Altarelli:2010gt,Ishimori:2010au,King:2013eh,King:2014nza,King:2015aea,King:2017guk,Feruglio:2019ktm}, the structure of the driving superpotential $\mathcal{W}_{driv}$ may be enforced by a $U(1)_R$ symmetry, with the driving superfield $\chi^0$ having $R=2$, the weighton $\phi$ and Higgs superfields having $R=0$ and the matter superfields having $R=1$, which prevents other superpotential terms appearing\footnote{At the low energy scale, after the inclusion of SUSY breaking effects, the $U(1)_R$ symmetry will be broken to the usual discrete R-parity~\cite{Altarelli:2010gt}. Such SUSY breaking effects may also modify the predictions from modular symmetry~\cite{Feruglio:2017spp}. However the study of SUSY breaking is beyond the scope of this paper.}. The $F$-flatness condition gives
\begin{equation}
\frac{\partial \mathcal{W}_{driv}}{\partial \chi^0}=Y^{(4)}_{\mathbf{1}}\frac{\phi^4}{M_{fl}^2} -M^2=0\,,
\end{equation}
which leads to the following VEV of the weighton $\phi$,
\begin{equation}
\langle\phi\rangle=\left(M^2M_{fl}^2/Y^{(4)}_{\mathbf{1}}\right)^{1/4}\,.
\end{equation}
After the weighton $\phi$ develops a VEV, the non-renormalisable terms are suppressed by powers of
\begin{equation}
\tilde{\phi}\equiv \frac{\langle \phi \rangle}{M_{fl}}\sim
\left(\frac{M}{M_{fl}}\right)^{1/2}\,,
\label{phitilde}
\end{equation}
where $M_{fl}$ is a dimensionful cut-off flavour scale.

\subsection{Yukawa Matrices}
\label{ChargedFermionSector}
In this subsection, we consider the leading order Yukawa operators, allowed by modular symmetry. There are no flavons, but we shall assume that there are several moduli which are located at different fixed points, as discussed earlier. As shown in~\cite{Baur:2019kwi,Baur:2019iai,Nilles:2020nnc}, different residual symmetries are preserved at different points in multi-dimensional moduli space such that fields which live at different locations in moduli space feel a different amount of modular symmetry. However, constructing a model with four different moduli in the up-type quarks, down-type quarks and neutrino sectors is out of the scope of the present work, and our approach is purely phenomenological here. Furthermore concrete models with several moduli frozen at distinct fixed points could be constructed, and certain flavons which are bi-triplets of the multiple finite modular groups are generally necessary~\cite{deMedeirosVarzielas:2019cyj,King:2019vhv,King:2021fhl}. For the up type quarks we write down the allowed non-renormalisable operators, and allow all possible group contractions, making use of the weighton field to generate mass hierarchies of the up and charm quarks as compared to the top quark mass which appears at the renormalisable level. For the down type quarks and charged leptons, the weighton field can also help to generate the mass hierarchies, and a mild fine-tuning is necessary to obtain the measured Cabibbo angle. There is a texture zero in the (1,1) element such that the GST relation is approximately satisfied. The neutrino masses and mixing arise from having two right-handed neutrinos with CSD(3.45).

\subsubsection{\label{subsubsec:up-quark} Up-type quarks}

The Yukawa matrix of the up-type quarks can be constructed by considering the non-renormalisable operators $TTH_5$, $T_3TH_5$ and the renormalisable operator $T_3T_3H_5$ for the (33) element. The non-renormalisable operators are suppressed by powers of a common mass scale $M_{fl}$ leading to powers of the flavor factor $\tilde{\phi}$ as in Eq.~\eqref{phitilde}. We assume that the modular symmetry is broken down to the $Z_3$ subgroup generated by $TS$. We shall use the modular forms at the fixed point $\tau_{TS}=-\omega^2$, namely the weight 2 modular form $Y^{(2)}_{\mathbf{2}}=Y_{TS} (1, 0)^T$, the weight 4 modular form $Y^{(4)}_{\mathbf{2}}=Y_{TS}^2 (0, 1)^T$, and the weight 6 modular form $Y^{(6)}_{\mathbf{1'}}=-Y_{ST}^{3}$, as shown in table~\ref{tab:residual-fixed-points}. The most important operators which generate a contribution to the up Yukawa matrix are\footnote{The term $(TT)_{\mathbf{1'}} H_5$ is exactly vanishing because the $S_4$ contraction rule for $\mathbf{2}\otimes\mathbf{2}\rightarrow \mathbf{1'}$ implies $(TT)_{\mathbf{1'}}=(0,0)^{T}$. Although the term $\tilde{\phi}^6Y^{(6)}_{\mathbf{1'}}(TT)_{\mathbf{1}} H_5$ would lead to non-vanishing (12) and (21) entries of up-type quark mass matrix, its contribution is suppressed by $\tilde{\phi}^6$.}
\begin{equation}
\alpha_u {\tilde{\phi}}^4 Y^{(4)}_{\mathbf{2}} (TT)_{\mathbf{2}} H_5+ \beta_u  {\tilde{\phi}}^2  Y^{(2)}_{\mathbf{2}} (TT)_{\mathbf{2}} H_5+
\gamma_uY^{(6)}_{\mathbf{1'}}T_3 T_3 H_5+\epsilon_u \tilde{\phi}T_3(TY^{(4)}_{\mathbf{2}})_{\mathbf{1'}}H_5\,.
\label{YuOperators_mod}
\end{equation}
Note that the lowest non-trivial modular weight containing the singlet is weight 4 and that is zero at the fixed point $Y^{(4)}_{\mathbf{1}}(\tau_{TS})=0$. This implies that the doublet contraction $(TT)_{\mathbf{2}}$ plays an important role, leading to a diagonal up type Yukawa matrix, using the Clebsch-Gordan coefficients for $\mathbf{2}\otimes\mathbf{2}$ in the Appendix~\ref{sec:S4_group_app}, and the fixed point values of the modular forms above, we have
\begin{equation}
\label{eq:Yu-origin}\mathcal{Y}^u_{\text{GUT}}\approx\left(
\begin{array}{ccc}
 \alpha_u {\tilde{\phi}}^4 & 0 & 0 \\
 0 &  \beta_u {\tilde{\phi}}^2 & \epsilon_u\tilde{\phi} \\
 0   &  \epsilon_u\tilde{\phi} &  \gamma_u
\end{array}
\right)\,,
\end{equation}
where the factors $Y^2_{TS}$, $Y_{TS}$, $Y^{(6)}_{\mathbf{1'}}$ and $Y^2_{TS}$ have been absorbed into the coupling constants $\alpha_u$, $\beta_u$, $\gamma_u$ and $\epsilon_u$ respectively. The parameters $\beta_u$ and $\gamma_u$ can be taken real by exploiting field redefinitions of $T$ and $T_3$. Moreover, the phase of $\alpha_u$ is irrelevant to both quark masses and CKM mixing matrix, and its phase can be absorbed into the right-handed charm quark. However, $\epsilon_u$ is generally a complex parameter. The suppression factor in Eq.~\eqref{phitilde} generates the up and charm quark mass hierarchy naturally, with $m_{u,c,t }\propto \tilde{\phi}^4, \tilde{\phi}^2, 1$, assuming $\alpha_u \sim \beta_u \sim \gamma_u\sim\mathcal{O}(1)$. It is well-known that mass hierarchy among the up quarks is $m_u:m_c:m_t\simeq\lambda^8:\lambda^4:1$ with $\lambda\simeq0.22$ being the Wolfenstein parameter. As a consequence, the weighton VEV $\tilde{\phi}$ is expected to be of order $\lambda^2$.

\subsubsection{\label{subsec:down-charged}Down-type quarks and charged leptons}

We assign the three generations of the matter fields $F$ to an $S_4$ triplet $\mathbf{3}$, the first two generations of the 10-plet transform as a doublet $\mathbf{2}$ under $S_4$. Thus there are two options: $(T_1, T_2)^{T}\sim\mathbf{2}$ and $(T_2, T_1)^{T}\sim\mathbf{2}$.

\begin{itemize}
\item{$(T_1,T_2)^{T}\sim\mathbf{2}$}\\
We have the following contractions for $T$ and $F$:
\begin{equation}
(TF)_{\mathbf{3}}\sim
\begin{pmatrix}
T_1F_2+T_2F_3 \\
T_1F_3+T_2F_1 \\
T_1F_1+T_2F_2
\end{pmatrix},~~~~(TF)_{\mathbf{3'}}\sim
\begin{pmatrix}
T_1F_2-T_2F_3 \\
T_1F_3-T_2F_1 \\
T_1F_1-T_2F_2
\end{pmatrix}
\end{equation}
We see that $T_1F_1$ and $T_2F_2$, which are related to the down and strange quark masses respectively, appear simultaneously as the third component of both contractions $(TF)_{\mathbf{3}}$ and $(TF)_{\mathbf{3'}}$. The operators $TFH_{\overline{\mathbf{5}}}$ and $TFH_{\overline{\mathbf{45}}}$ combining with modular form $Y_{\mathbf{3}}$ or $Y_{\mathbf{3'}}$, generate the masses of the down quarks and charged leptons. As a result, the down and strange quark messes would be of the same order except for the case that the contributions of $(TF)_{\mathbf{3}}$ and $(TF)_{\mathbf{3'}}$ cancel with each other.
\item{$(T_2,T_1)^{T}\sim\mathbf{2}$}\\
We have the following contractions for $T$ and $F$:
\begin{equation}
(TF)_{\mathbf{3}}\sim
\begin{pmatrix}
T_2F_2+T_1F_3 \\
T_2F_3+T_1F_1 \\
T_2F_1+T_1F_2
\end{pmatrix},~~~~(TF)_{\mathbf{3'}}\sim
\begin{pmatrix}
T_2F_2-T_1F_3 \\
T_2F_3-T_1F_1 \\
T_2F_1-T_1F_2
\end{pmatrix}
\end{equation}
We see that $T_2F_2$ and $T_1F_1$ appear in the 1st and 2nd components of these contractions respectively. Hence fine-tuning is not necessary to explain the mass hierarchies between the down and strange quarks. This is the reason why the assignment $(T_2,T_1)^{T}\sim\mathbf{2}$ is chosen in the model~\cite{Ding:2010pc}.
\end{itemize}
Then the Yukawa matrices of the down-type quarks and the charged leptons can be deduced from the leading superpotential operators\footnote{Here $Y^{(6)}_{\mathbf{3}}$ refers to $Y^{(6)}_{\mathbf{3}, II}$ since $Y^{(6)}_{\mathbf{3}, I}$ is vanishing with $Y^{(6)}_{\mathbf{3}, I}=(0, 0, 0)$ at the fixed point $\tau=\tau_{ST}$, as shown in table~\ref{tab:residual-fixed-points}. Moreover, $Y^{(8)}_{\mathbf{3}}$ and $Y^{(8)}_{\mathbf{3'}}$ stand for the two independent weight 8 triplet modular forms transforming as $\mathbf{3}$ and $\mathbf{3'}$ respectively, and they are proportional to $(0, 1, 0)$ at residual modular symmetry conserving point $\tau_{ST}$. $Y^{(10)}_{\mathbf{3}}$ denotes the three independent weight 10 modular forms in the representation $\mathbf{3}$, and its values is proportional to $(0, 0, 1)$ at the fixed point $\tau=\tau_{ST}$. }
\begin{eqnarray}
\nonumber&&\alpha_{d1}   {\tilde{\phi}}^3(Y^{(8)}_{\mathbf{3}} (TF)_{\mathbf{3}})_{\mathbf{1}}  H_{\overline{5}}
+\alpha_{d2}   {\tilde{\phi}}^3(Y^{(8)}_{\mathbf{3'}} (TF)_{\mathbf{3'}})_{\mathbf{1}}  H_{\overline{5}}+
\beta_{d1}   {\tilde{\phi}}(Y^{(6)}_{\mathbf{3}}
(TF)_{\mathbf{3}})_{\mathbf{1}}  H_{\overline{5}}\\
\nonumber&&+\beta_{d2}   {\tilde{\phi}}(Y^{(6)}_{\mathbf{3'}} (TF)_{\mathbf{3'}})_{\mathbf{1}}  H_{\overline{5}}+
\gamma_d (Y^{(8)}_{\mathbf{3}} F  )_{\mathbf{1}} T_3 H_{\overline{5}}+\epsilon_d  {\tilde{\phi}}^2(Y^{(10)}_{\mathbf{3}} F  )_{\mathbf{1}} T_3 H_{\overline{5}}
\\
\nonumber&&+
\alpha'_{d1}   {\tilde{\phi}}^3(Y^{(8)}_{\mathbf{3}} (TF)_{\mathbf{3}})_{\mathbf{1}}  H_{\overline{45}}
+\alpha'_{d2}   {\tilde{\phi}}^3(Y^{(8)}_{\mathbf{3'}} (TF)_{\mathbf{3'}})_{\mathbf{1}}  H_{\overline{45}}+
\beta'_{d1}   {\tilde{\phi}}(Y^{(6)}_{\mathbf{3}}
(TF)_{\mathbf{3}})_{\mathbf{1}}  H_{\overline{45}}\\
&&+\beta'_{d2}   {\tilde{\phi}}(Y^{(6)}_{\mathbf{3'}} (TF)_{\mathbf{3'}})_{\mathbf{1}}  H_{\overline{45}}+
\gamma'_d (Y^{(8)}_{\mathbf{3}} F  )_{\mathbf{1}} T_3 H_{\overline{45}}+\epsilon'_d  {\tilde{\phi}}^2(Y^{(10)}_{\mathbf{3}} F  )_{\mathbf{1}} T_3 H_{\overline{45}}\,.
\label{YdOperators_mod-va-4th}
\end{eqnarray}
Notice that the Yukawa couplings to $H_{\overline{45}}$ replicate those of $H_{\overline{5}}$ because both Higgs multiplets $H_{\overline{5}}$ and $H_{\overline{45}}$ are $S_4$ singlets with the same modular weight. The phase of the coupling $\gamma_d$ is irrelevant and it can be absorbed by redefinition of the supermultiplet $F$. The modular symmetry is assumed to be spontaneously broken down to a $Z_3$ subgroup generated by $ST$, and the alignments of the modular forms are $Y^{(6)}_{\mathbf{3}}=Y^{(6)}_{\mathbf{3'}}=\sqrt{3}\omega Y_{ST}^{3} (1,0,0)$, $Y^{(8)}_{\mathbf{3}, \mathbf{3'}}\propto(0, 1, 0)$ and $Y^{(10)}_{\mathbf{3}}\propto(0, 0, 1)$ at the fixed point $\tau_{ST}$, as shown in table~\ref{tab:residual-fixed-points}.  Using the Clebsch-Gordan coefficients for the different $S_4$ contractions as in the Appendix~\ref{sec:S4_group_app} and separating the contributions of $H_{\overline{5}}$ and $H_{\overline{45}}$, we find
\begin{eqnarray}
\nonumber&&\mathcal Y_{\overline{5}}\approx\left(
\begin{array}{ccc}
0 & (\alpha_{d1}+\alpha_{d2})\tilde{\phi}^3  & 0 \\
 (\alpha_{d1}-\alpha_{d2})\tilde{\phi}^3 & (\beta_{d1}+\beta_{d2})\tilde{\phi}  & \epsilon_d\tilde{\phi}^2  \\
(\beta_{d1}-\beta_{d2})\tilde{\phi} & 0 & \gamma_d
\end{array}
\right) \,,\\
\label{eq:Y5bar-45bar}&& \mathcal Y_{\overline{45}}\approx\left(
\begin{array}{ccc}
0 & (\alpha'_{d1}+\alpha'_{d2})\tilde{\phi}^3  & 0 \\
 (\alpha'_{d1}-\alpha'_{d2})\tilde{\phi}^3 & (\beta'_{d1}+\beta'_{d2})\tilde{\phi}  & \epsilon'_d\tilde{\phi}^2  \\
(\beta'_{d1}-\beta'_{d2})\tilde{\phi} & 0 & \gamma'_d
\end{array}
\right) \,,
\end{eqnarray}
where the convention for the above Yukawa coupling is $F_i(\mathcal{Y}_{\overline{5}})_{ij}T_j$ and $F_i(\mathcal{Y}_{\overline{45}})_{ij}T_j$.  Notice that the (11) and (32) elements of $\mathcal{Y}_{\overline{5}}$ and $\mathcal{Y}_{\overline{45}}$ can arise from the operators ${\tilde{\phi}}^5(Y^{(10)}_{\mathbf{3}} (TF)_{\mathbf{3}})_{\mathbf{1}} H_{\overline{5}}$, ${\tilde{\phi}}^5(Y^{(10)}_{\mathbf{3'}} (TF)_{\mathbf{3'}})_{\mathbf{1}} H_{\overline{5}}$, ${\tilde{\phi}}^5(Y^{(10)}_{\mathbf{3}} (TF)_{\mathbf{3}})_{\mathbf{1}} H_{\overline{45}}$ and ${\tilde{\phi}}^5(Y^{(10)}_{\mathbf{3'}} (TF)_{\mathbf{3'}})_{\mathbf{1}} H_{\overline{45}}$, and they are suppressed by ${\tilde{\phi}}^5$. Similarly ${\tilde{\phi}}^4(Y^{(12)}_{\mathbf{3}} F )_{\mathbf{1}} T_3 H_{\overline{5}}$ and ${\tilde{\phi}}^4(Y^{(12)}_{\mathbf{3}} F)_{\mathbf{1}} T_3 H_{\overline{45}}$ can lead to non-vanishing (13) entry. The Yukawa matrices of the down-type quarks and the charged leptons are linear combinations of the two structures in Eq.~\eqref{eq:Y5bar-45bar}. Following the construction proposed by Georgi and Jarlskog~\cite{Georgi:1979df}, we have
\be
\label{Ydb-modular}
\mathcal Y^e_{\text{GUT}}=\mathcal Y_{\overline{5}}-3\mathcal Y_{\overline{45}}, \ \ \ \
\mathcal Y^d_{\text{GUT}}=(\mathcal Y_{\overline{5}}+\mathcal Y_{\overline{45}})^T \,.
\ee
Thus the charged lepton and down quark mass matrices are given by $M_e=\mathcal{Y}^e_{\text{GUT}} v_d$ and $M_d=\mathcal{Y}^d_{\text{GUT}} v_d$ respectively. The phase of $\gamma_d$ can be removed by the field redefinition of the $\overline{\mathbf{5}}$ matter field $F$ while all the other parameters are complex. Since the (11) element of the down quark Yukawa coupling matrix is vanishing, the Gatto-Sartori-Tonin (GST) relation $\theta^{q}_{12}\simeq\sqrt{m_d/m_s}$~\cite{Gatto:1968ss} is approximately satisfied. However, the (12) and (21) entries are suppressed by $\tilde{\phi}^3$ so that the Cabibbo angle is expected to be of order $\tilde{\phi}^2$ if all coupling constants are of order one. Thus the parameters $\alpha_{d1,d2}$ and $\alpha'_{d1,d2}$ should be relatively large to reproduce the correct size of Cabibbo angle. This point is confirmed in the numerical calculation, as shown in section~\ref{numerical}.

\subsubsection{\label{sec:tridirec_MM}Neutrino Mass and Mixing}

In the neutrino sector we have two right-handed neutrinos in $S_4$ representations $N_a\sim \mathbf{1'}$ and $N_s\sim \mathbf{1}$, where the respective Dirac Yukawa couplings are determined by the fixed points from Eq.~\eqref{2+i} and table~\ref{tab:residual-fixed-points}:
\begin{eqnarray}
Y^{(6)}_\mathbf{3'}\propto \begin{pmatrix}
0  \\
1  \\
-1 \\
\end{pmatrix}, ~~~Y^{(2)}_{\mathbf{3}} \propto\begin{pmatrix}
1  \\
1+\sqrt{6}  \\
1-\sqrt{6} \\
\end{pmatrix}\,
\end{eqnarray}
We note that, in the CSD($n$) model, the two columns of the Dirac mass matrix are proportional to $\left(0, 1, -1\right)$ and $\left(1, n, 2-n\right)$ respectively, \cite{King:2013iva,King:2013xba,King:2015dvf,Chen:2019oey}, so this corresponds approximately to the case CSD($3.45$)~\cite{Gui-JunDing:2019wap}.  By comparison, the predictive Littlest Seesaw model and its variant are the cases of $n=3$~\cite{King:2013iva,King:2015dvf,King:2016yvg,Ballett:2016yod,King:2018fqh}, $n=4$~\cite{King:2013xba,King:2013hoa,King:2014iia,Bjorkeroth:2014vha} and $n=-1/2$~\cite{Chen:2019oey} respectively. It has been shown that the CSD($n$) model can be reproduced from the $S_4$ flavour symmetry in the tri-direct CP approach~\cite{Ding:2018fyz,Ding:2018tuj}, where the parameter $n$ is constrained to be a generic real parameter by the $S_4$ flavour symmetry and CP symmetry~\cite{Ding:2018fyz,Ding:2018tuj}. Here the modular symmetry can fix the alignment parameter $n$ to be $1+\sqrt{6}\approx 3.45$. This is a remarkable advantage of modular symmetry with respect to discrete flavour symmetry.

The most important operators for the neutrino masses are
\begin{eqnarray}
\nonumber && \alpha_{\nu}{\tilde{\phi}} (Y^{(6)}_{\mathbf{3'}} F )_{\mathbf{1'}} N_{\mathrm{a}} H_{{5}}
+
\beta_{\nu}{\tilde{\phi}}^2 (Y^{(2)}_{\mathbf{3}} F )_{\mathbf{1}} N_{\mathrm{s}} H_{{5}}
\\
\label{eq:TD_MM_Lag}&&
-\frac{1}{2}M^{(8)}_{\mathbf{1}}Y^{(8)}_{\mathbf{1}}N_{\mathrm{a}}N_{\mathrm{a}}
-\frac{1}{2}M^{(0)}_{\mathbf{1}}N_{\mathrm{s}}N_{\mathrm{s}}{\tilde{\phi}}^2
-M^{(6)}_{\mathbf{1'}}Y^{(6)}_{\mathbf{1'}}N_{\mathrm{a}}N_{\mathrm{s}}{\tilde{\phi}}^3\,.
\end{eqnarray}
The neutrino Dirac mass matrix and the Majorana mass matrix of right-handed neutrinos are,
\begin{equation}
m_{D}=\begin{pmatrix}
\beta_{\nu}{\tilde{\phi}}^2 Y^{(2)}_{\mathbf{3}} v_u,  &
\alpha_{\nu}{\tilde{\phi}} Y^{(6)}_{\mathbf{3'}}v_u
\end{pmatrix},\qquad
m_{N}=\begin{pmatrix}
M^{(0)}_{\mathbf{1}}{\tilde{\phi}}^2 ~&~ M^{(6)}_{\mathbf{1'}}{\tilde{\phi}}^3 \\
M^{(6)}_{\mathbf{1'}}{\tilde{\phi}}^3 ~& ~
M^{(8)}_{\mathbf{1}}
\end{pmatrix}\,,
\end{equation}
where the Clebsch-Gordan coefficients in both contractions are omitted for notation simplicity, $Y^{(8)}_{\mathbf{1}}$ and $Y^{(6)}_{\mathbf{1'}}$ are absorbed into $M^{(8)}_{\mathbf{1}}$ and $M^{(6)}_{\mathbf{1'}}$ respectively. The heavy Majorana mass matrix is approximately diagonal to excellent approximation. The effective light neutrino mass matrix is given by the seesaw formula $m_{\nu}=-m_Dm_N^{-1}m^T_D$ which implies
\begin{equation}
\label{eq:mnu_TDM}
m_{\nu}=
- \frac{\alpha_{\nu}^2{\tilde{\phi}}^2v^2_u}{M^{(8)}_{\mathbf{1}} }Y^{(6)}_{\mathbf{3'}}{Y^{(6)}_{\mathbf{3'}}}^T
- \frac{\beta_{\nu}^2{\tilde{\phi}}^2v^2_u}{M^{(0)}_{\mathbf{1}} }Y^{(2)}_{\mathbf{3}}{Y^{(2)}_{\mathbf{3}}}^T\,,
\end{equation}
where the two terms are equally suppressed by ${\tilde{\phi}}^2$. From Eq.~\eqref{eq:mnu_TDM}, we find the neutrino mass matrix is predicted to be
\begin{equation}
\label{eq:mnu-GUT} m_{\nu}=m_a\left(
\begin{array}{ccc}
 0 & 0 & 0 \\
 0 & 1 & -1 \\
 0 & -1 & 1 \\
\end{array}
\right)+m_se^{i\eta}\left(
\begin{array}{ccc}
 1 & 1-\sqrt{6} & 1+\sqrt{6} \\
 1-\sqrt{6} & 7-2 \sqrt{6} & -5 \\
 1+\sqrt{6} & -5 & 7+2 \sqrt{6} \\
\end{array}
\right)\,.
\end{equation}
It is notable that only three free parameters $m_a$, $m_s$ and $\eta$ are involved in the neutrino mass matrix. It is straightforward to check that the column vector $\left(2, -1, -1\right)^{T}$ is an eigenvector of $m_{\nu}$ with vanishing eigenvalue. Therefore the neutrino mass spectrum is normal ordering, and lightest neutrino is massless $m_1=0$, and the neutrino mixing matrix is determined to be the TM$_1$ pattern,
\begin{equation}
U_{\nu}=\begin{pmatrix}
\sqrt{\frac{2}{3}}  &  -  &  -  \\
-\frac{1}{\sqrt{6}}  &  -  &  -   \\
-\frac{1}{\sqrt{6}}  &  -  &  -
\end{pmatrix}\,.
\end{equation}
The charged lepton mass matrix given by Eq.~\eqref{Ydb-modular} is not diagonal, consequently we have to include the charged lepton corrections to the lepton mixing matrix.  Using the general formulae for neutrino masses and mixing angles for the Littlest Seesaw~\cite{King:2015dvf} and tri-direct model~\cite{Ding:2018fyz,Ding:2018tuj}, we see that the experimental data can be accommodated very well for certain values of $m_a$, $m_s$ and $\eta$, as discussed in next section.

\section{Numerical analysis }
\label{numerical}

In this section, we shall discuss whether the model is compatible with the experimental data through a detailed numerical analysis, and we take the VEVs of weighton as $\tilde{\phi}=0.1$ for illustration. The Yukawa matrices in Eqs.~(\ref{eq:Yu-origin}, \ref{Ydb-modular}) and the light neutrino mass matrix in Eq.~\eqref{eq:mnu-GUT} are predicted  at the GUT scale, thus we should compare them with the data of fermion masses and mixing parameters obtained by performing renormalization group evolution of their measured values. As regards the charged fermion masses and the quark mixing parameters, we take a representative set of data at the GUT scale from~\cite{Antusch:2013jca} for $\tan\beta=10$ and the SUSY breaking scale $M_{\text{SUSY}}=10$ TeV. Regarding the neutrino masses and mixing angles, we use the values of the latest global fit NuFIT 5.0~\cite{Esteban:2020cvm}, and thus we neglect the effect of renormalization group evolution from low energy scale to the GUT scale. Since two right-handed neutrinos are introduced in our model, the lightest neutrino is massless. It is known that the running of neutrino masses and mixing parameters is negligible for strongly hierarchical neutrino spectrum.

As discussed in section~\ref{ChargedFermionSector}, the up type quark Yukawa couplings have three real parameters $\alpha_u$, $\beta_u$, $\gamma_u$ and one complex parameter $\epsilon_u$, the down quark and charged lepton sectors involve 12 parameters $\alpha_{d1, d2}$, $\beta_{d1, d2}$, $\gamma_d$, $\epsilon_d$, $\alpha'_{d1, d2}$, $\beta'_{d1, d2}$, $\gamma'_d$, $\epsilon'_d$ which are complex numbers except $\gamma_d$, and light neutrino mass matrix only depends on three real parameters $m_a$, $m_s$ and $\eta$. We optimize the values of these parameters by using the conventional $\chi^2$ analysis, and the $\chi^2$ function is defined as
\begin{equation}
\chi^2=\sum^{n}_{i=1}\left(\frac{P_i-O_i}{\sigma_i}\right)^2\,,
\end{equation}
where $O_i$ and $\sigma_i$ denote the central values and the $1\sigma$ errors of the corresponding quantities shown in table~\ref{Tab:fitted_fermion_parameters}, and $P_i$ are the predictions for the corresponding observables as complex functions of free parameters of the models. We numerically diagonalize the Yukawa matrices $\mathcal{Y}^u_{\text{GUT}}$, $\mathcal{Y}^e_{\text{GUT}}$, $\mathcal{Y}^d_{\text{GUT}}$ as well as the light neutrino mass matrix $m_{\nu}$, then we can obtain the predictions for the masses, mixing angles and CP violation phases of both quarks and leptons. The lepton mixing matrix is parameterized in terms of mixing angles and CP phases in the convention of PDG parameterization~\cite{Zyla:2020zbs},
\begin{eqnarray}
 \label{eq:UmatrixPDG}
U =\left(\begin{array}{ccc}
 c_{12} c_{13} & s_{12} c_{13}   & s_{13} e^{-i\delta^l_{CP}}  \\
 - s_{12} c_{23} - c_{12} s_{13} s_{23} e^{i\delta^l_{CP}}  & \hphantom{+} c_{12} c_{23} - s_{12} s_{13} s_{23} e^{i\delta^l_{CP}}  & c_{13} s_{23} \hspace*{5.5mm}\\
\hphantom{+} s_{12} s_{23} - c_{12} s_{13} c_{23} e^{i\delta^l_{CP}} & - c_{12} s_{23} - s_{12} s_{13} c_{23} e^{i\delta^l_{CP}} & c_{13} c_{23} \hspace*{5.5mm}
\end{array}\right)
\left(\begin{array}{ccc}
e^{i\frac{\rho_1}{2}} & 0 & 0 \\
0 & e^{i\frac{\rho_2}{2}} & 0\\
0 & 0 & 1 \\
\end{array}\right)\,,
\end{eqnarray}
where $c_{ij}\equiv\cos \theta^l_{ij}, s_{ij}\equiv\sin\theta^l_{ij}, \delta^l_{CP}$ is the Dirac CP phase and $\rho_{1, 2}$ are Majorana CP phases. In our model the Majorana CP phase $\rho_1$ is unphysical since $m_1=0$ therefore the only physical Majorana CP phase is $\rho\equiv\rho_2$. The quark mixing matrix is parameterized in a similar way without Majorana CP phases. The ratios between the charged fermion masses are included in the $\chi^2$ function, and the measured masses of the third generation can be exactly reproduced by adjusting the overall scales of the mass matrices. Similarly we include the ratio $\Delta m^2_{21}/\Delta m^2_{31}$ instead of $\Delta m^2_{21}$ and $\Delta m^2_{31}$ individually in $\chi^2$. The absolute values of the ratios of the Yukawa couplings are treated as random numbers between 0 and 1000, and the phases of complex couplings vary in the range of $[0, 2\pi]$. The numerical minimization of the $\chi^2$ function gives the best fit point in parameter space which minimizes the $\chi^2$, as shown in table~\ref{tab:best-fit-input}. From the fitted values of the parameters, one can obtain the predictions for the masses and mixing parameters in the quark and lepton sectors. The results of the fit are summarized in table~\ref{Tab:fitted_fermion_parameters}. Obviously the model can accommodate the experimental data very well, all the observables fall within the $1\sigma$ experimental ranges.

\begin{table}[htpb!]
\begin{small}
  \begin{tabular}{|c|c|c|c|c|c|c|c|}
    \hline\hline
$\alpha_{u}/\gamma_{u}$ & $0.05451$ & $\beta_{u}/\gamma_{u}$ & $0.4438$ & $|\epsilon_{u}/\gamma_{u}|$ & $0.4009$ & $\text{arg}(\epsilon_{u}/\gamma_{u})$ & $1.9999\pi$ \\ \hline
$|\alpha_{d1}/\gamma_{d}|$ & $54.7737$ & $\text{arg}(\alpha_{d1}/\gamma_{d})$ & $0.05954\pi$ &
$|\alpha_{d2}/\gamma_{d}|$ & $61.1089$ & $\text{arg}(\alpha_{d2}/\gamma_{d})$ & $1.01548\pi$ \\ \hline
$|\beta_{d1}/\gamma_{d}|$ & $2.2516$ & $\text{arg}(\beta_{d1}/\gamma_{d})$ & $0.004161\pi$ &
$|\beta_{d2}/\gamma_{d}|$ & $2.1808$ & $\text{arg}(\beta_{d2}/\gamma_{d})$ & $0.9956\pi$ \\ \hline
$|\alpha'_{d1}/\gamma_{d}|$ & $54.4400$ & $\text{arg}(\alpha'_{d1}/\gamma_{d})$ & $1.05652\pi$ &
$|\alpha'_{d2}/\gamma_{d}|$ & $55.7477$ & $\text{arg}(\alpha'_{d2}/\gamma_{d})$ & $0.02208\pi$ \\ \hline
$|\beta'_{d1}/\gamma_{d}|$ & $2.2924$ & $\text{arg}(\beta'_{d1}/\gamma_{d})$ & $1.01751\pi$ &
$|\beta'_{d2}/\gamma_{d}|$ & $2.1003$ & $\text{arg}(\beta'_{d2}/\gamma_{d})$ & $1.9799\pi$ \\ \hline
$|\epsilon_{d}/\gamma_{d}|$ & $0.5470$ & $\text{arg}(\epsilon_{d}/\gamma_{d})$ & $1.05092\pi$ &$|\epsilon'_{d}/\gamma_{d}|$ & $0.1625$ & $\text{arg}(\epsilon'_{d}/\gamma_{d})$ & $1.9962\pi$ \\ \hline
$ \gamma'_{d}/\gamma_{d}$ & $0.3494$ & $m_{s}/m_{a}$ & $0.1374$ &$\eta$ & $1.3330\pi$ & \multicolumn{2}{c|}{$\tilde{\phi}=0.1$} \\ \hline\hline
\end{tabular}
\end{small}
\caption{\label{tab:best-fit-input} The best fit values of the input parameters at the minimum of the $\chi^2$.}
\end{table}

The overall mass scales $\gamma_{u}v_u$ and $\gamma_{d}v_d$ are fixed by the top quark and tau masses $m_t\simeq87.4555$ GeV, $m_{\tau}\simeq1.30234$ GeV respectively at GUT scale~\cite{Antusch:2013jca},
\begin{equation}
\gamma_{u}v_u=87.31479~\text{GeV},~~~~~\gamma_{d}v_d=0.71748~\text{GeV}\,.
\end{equation}
For $\tan\beta=10$, we can obtain the values of the Yukawa couplings $\gamma_u$ and $\gamma_d$ as follow
\begin{equation}
\gamma_u=0.504,~~~~\gamma_d=0.0414\,.
\end{equation}
The absolute values of other couplings can be extracted from their ratios with $\gamma_{u}$ and $\gamma_d$ given in table~\ref{tab:best-fit-input}.
Clearly some of the absolute dimensionless couplings such as $\gamma_d$
are not of order unity, as they would expected to be according to the weighton approach. Also we see that the couplings $\alpha_{d1,d2}$ and $\alpha'_{d1,d2}$ are large (but remain perturbative) in order to reproduce the observed values of the Cabibbo angle, as has been emphasized in section~\ref{subsec:down-charged}. However, the hierarchy among the coupling constants is much less than that of quark masses. As shown in Eq.~\eqref{Ydb-modular}, the charged lepton and down quark mass matrices are closely related and they depend on the same set of parameters.  Thus we can determine the bottom quark mass as
\begin{equation}
m_{b}\simeq0.96819~\text{GeV}\,.
\end{equation}
It is quite close to tau mass, consequently approximate $b-\tau$ unification is predicted in the model. The parameter $m_a$ is fixed by requiring that the mass squared difference $\Delta m^2_{21}=7.42\times 10^{-5} \text{eV}^2$ is reproduced,
\begin{equation}
m_{a}=22.896~\text{meV}\,.
\end{equation}
Thus the absolute values of the light neutrino masses can be determined,
\begin{equation}
m_1=0~\text{meV},~~~m_2=8.61394~\text{meV},~~~m_3=50.17078~\text{meV}\,.
\end{equation}
The neutrino mass spectrum is normal ordering, and the lightest neutrino is massless as only two right-handed neutrinos are introduced in the model. Moreover, the Majorana CP phase $\rho$ and the effective Majorana mass in neutrinoless double beta decay are found to be
\begin{equation}
\rho=17.68308^{\circ},~~~m_{\beta\beta}\simeq3.366~\text{meV}\,.
\end{equation}
In the present model, the lepton mixing matrix receives corrections from the charged lepton sector. At the best fitting point, the charged lepton diagonalization matrix is found to be
\begin{equation}
U_{l}=\left(
\begin{array}{ccc}
 -0.2763+0.9247 i & -0.0750+0.2510 i & 0.000063-0.00021i \\
 0.2520-0.0308i & -0.9284+0.1135 i & 0.2446-0.0299i   \\
 -0.0643 & 0.2379 & 0.9692 \\
\end{array}
\right)\,.
\end{equation}

\begin{table}[t!]
\centering
\begin{small}
\begin{tabular}{|c|c|c|}
\hline
\hline

observables
& best-fit $\pm$ $1\sigma$ & predictions \\ \hline

$\theta_{12}^{q}$[rad]
& $0.22736\pm 0.00073$
& $0.22736$ \\
\hline
$\theta_{13}^{q}$[rad]
& $0.00349\pm 0.00013$
& $0.00349$ \\
\hline
$\theta_{23}^{q}$[rad]
& $0.04015\pm 0.00064$
& $0.04015$ \\
\hline
$\delta^{q}_{CP}[^\circ]$
& $69.21330\pm 3.11460$
& $69.21632$ \\
\hline
$m_{u}/m_{c}$
& $(1.9286 \pm 0.6017) \times 10^{-3}$
& $1.9291 \times 10^{-3}$ \\
\hline
$m_{c}/m_{t}$
& $(2.8213 \pm 0.1195) \times 10^{-3}$
& $2.8213 \times 10^{-3}$ \\
\hline
$m_{d}/m_{s}$
& $(5.0523 \pm 0.6191) \times 10^{-2}$
& $5.0532 \times 10^{-2}$ \\
\hline
$m_{s}/m_{b}$
& $(1.8241 \pm 0.1005) \times 10^{-2}$
& $1.8240 \times 10^{-2}$ \\
  \hline
$m_{t}~[\text{GeV}]$ & $87.45553 \pm 2.08874$ & $87.45553$ \\
\hline
$m_{b}~[\text{GeV}]$ & $0.96819 \pm 0.01063$ & $0.96819$ \\
\hline
\hline
$\sin^2\theta_{12}^{l}$ (NO)
&$0.304_{-0.012}^{+0.012}$ & $0.30402$ \\
\hline
$\sin^2\theta_{13}^{l}$ (NO)
& $0.02219_{-0.00063}^{+0.00062}$ & $0.02219$ \\
\hline
$\sin^2\theta_{23}^{l}$ (NO)
& $0.573_{-0.020}^{+0.016}$ & $0.57302$ \\
\hline
$\delta_{CP}^{l}[^\circ]$ (NO)
& $197_{-24}^{+27}$ & $197.01325$ \\
\hline
$m_{e}/m_{\mu}$ & $(4.73689 \pm 0.04019) \times 10^{-3}$ & $4.73683 \times 10^{-3}$  \\
\hline
$m_{\mu}/m_{\tau}$ & $(5.85684 \pm 0.04654) \times 10^{-2}$ & $5.85684 \times 10^{-2}$  \\
\hline
$m_{\tau}~[\text{GeV}]$ & $1.30234 \pm 0.00679$ & $1.30234$ \\
  \hline
$\Delta m_{21}^{2}~[10^{-5}\text{eV}^{2}]$ (NO)
&  $7.42_{-0.20}^{+0.21}$ & $7.42000$ \\
\hline
$\Delta m_{31}^{2}~[10^{-3}\text{eV}^{2}]$ (NO)
&  $2.517_{-0.028}^{+0.026}$ & $2.51711$ \\
\hline
$\rho/^{\circ}$ (NO)
& $-$ & $17.68308$ \\
\hline
$m_1[\mathrm{meV}]$ (NO)
& $-$ & $0$ \\
\hline
$m_2[\mathrm{meV}]$ (NO)
& $-$ & $8.61394$ \\
\hline
$m_3[\mathrm{meV}]$ (NO)
& $-$ & $50.17078$ \\ \hline

$\chi^2_{\text{min}}$  &  &  $2.4057\times 10^{-5}$ \\
\hline
\hline
\end{tabular}
\end{small}
\caption{\label{Tab:fitted_fermion_parameters} The predicted values of the masses and mixing parameters of quark and lepton in the model.
The best fit values and $1\sigma$ uncertainties of the quark and lepton parameters are evolved to the GUT scale as calculated in~\cite{Antusch:2013jca}, with the SUSY breaking scale $M_{\text{SUSY}}=10$ TeV and $\tan\beta=10$.
The data of lepton mixing angles, Dirac CP violation phases
$\delta^{l}_{CP}$ and the neutrino mass squared differences are taken from
NuFIT 5.0~\cite{Esteban:2020cvm}. }
\end{table}

\begin{figure}[t!]
\centering
\includegraphics[width=0.99\linewidth]{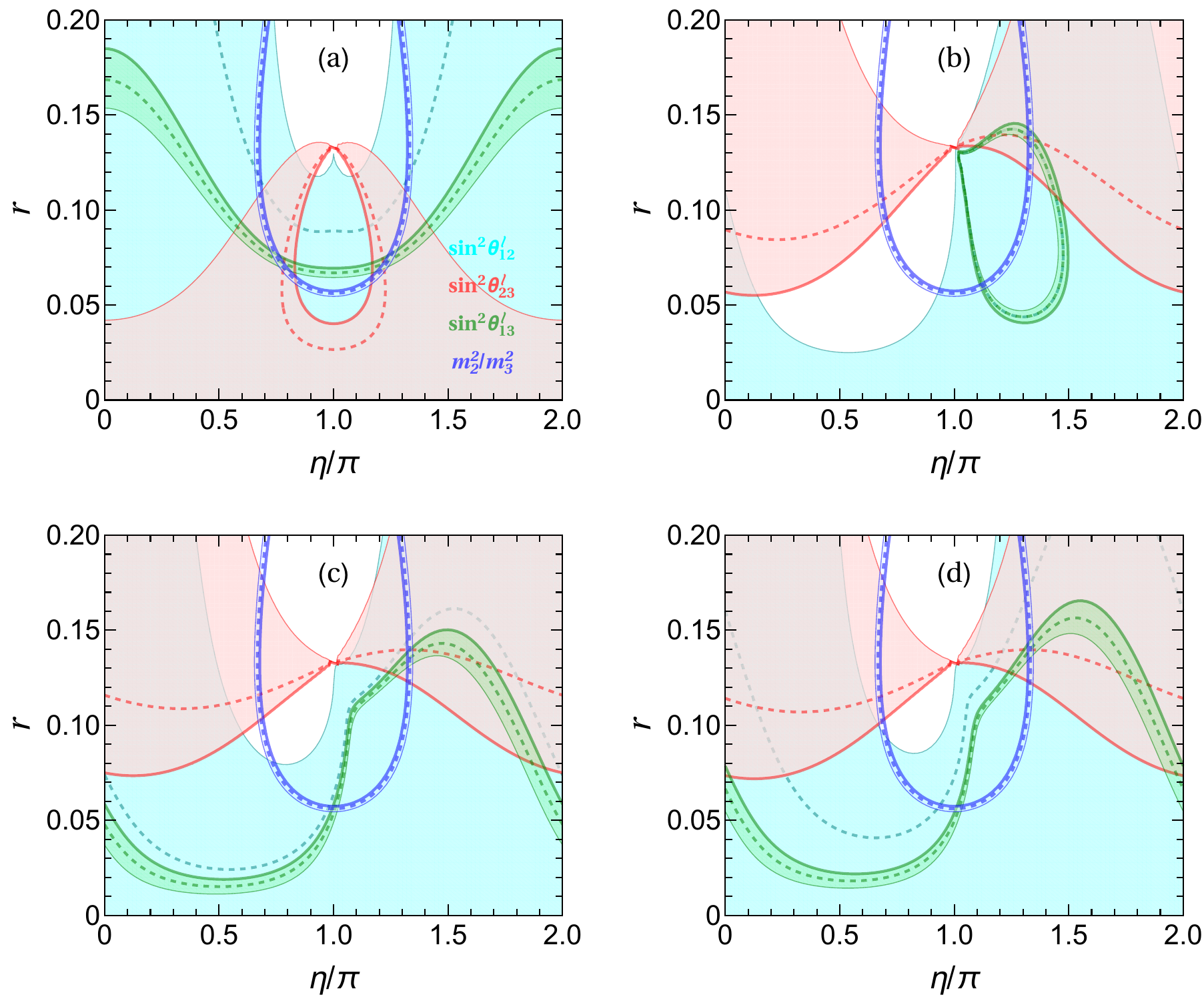}
\caption{\label{fig:contour}The contour plots of $\sin^2\theta^l_{12}$, $\sin^2\theta^l_{13}$, $\sin^2\theta^l_{23}$ and $m^2_2/m^2_3$ in the plane $r=m_s/m_a$ versus $\eta/\pi$. The cyan, red, green and blue areas denote the $3\sigma$ regions of $\sin^2\theta^l_{12}$, $\sin^2\theta^l_{13}$, $\sin^2\theta^l_{23}$ and $m^2_2/m^2_3$ respectively. The solid lines denote the 3 sigma upper bounds, the thin lines denote the 3 sigma lower bounds and the dashed lines refer to their best fit values~\cite{Esteban:2020cvm}. The panel (a) is for CSD(3.45) without charged lepton correction~\cite{Gui-JunDing:2019wap}. For the panel (b), the input parameters (except $m_s$, $m_a$ and $\eta$) are taken to be the best fit values shown in table~\ref{tab:best-fit-input}. The panels (c) and (d) show two very similar plots for another two local minima of $\chi^2$, namely the first and second local minima, respectively, discussed in Appendix~\ref{sec:app-local-minima}. }
\end{figure}

\begin{figure}[t!]
\centering
\includegraphics[width=1.0\linewidth]{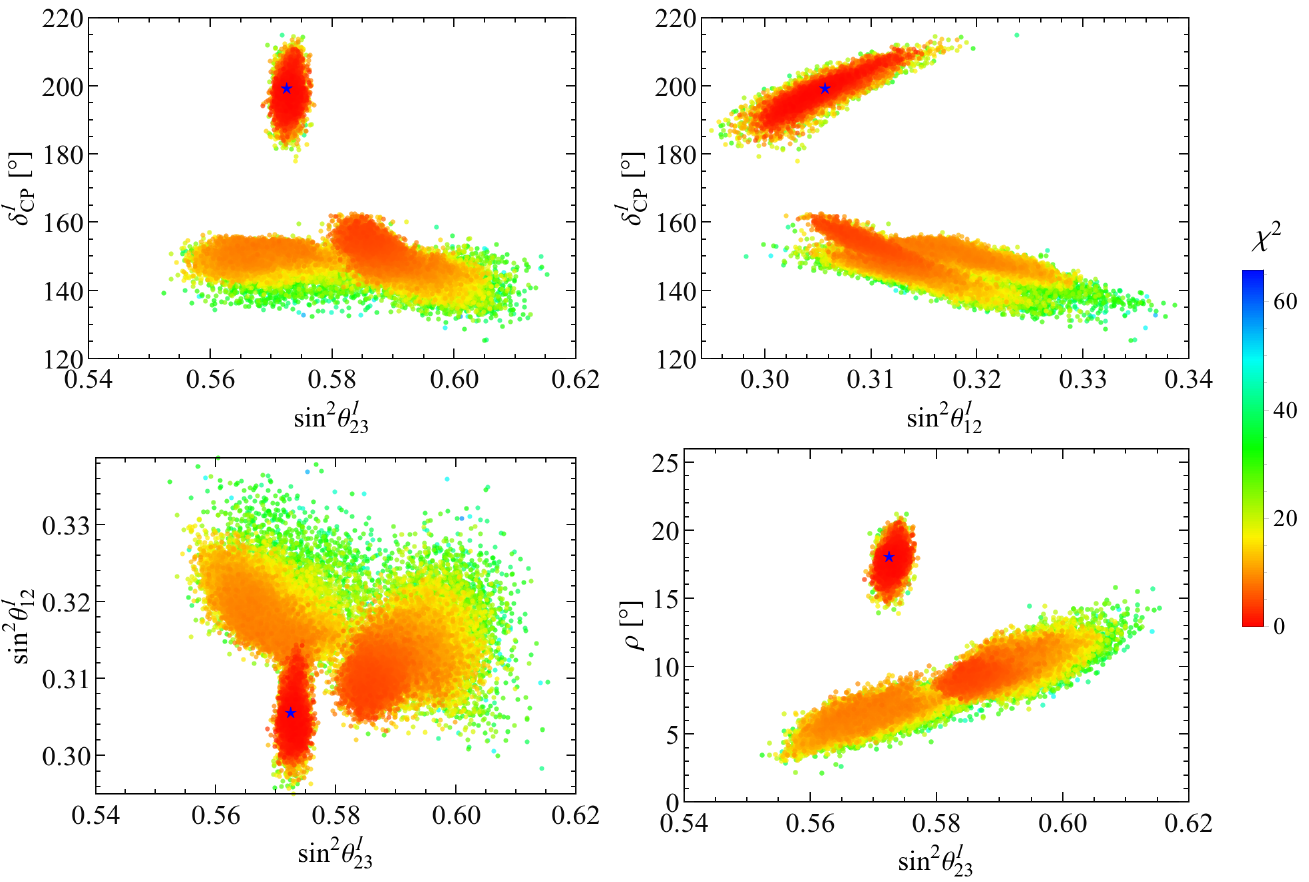}
\caption{\label{fig:correlations-lepton} Numerical results for the correlations among the lepton mixing parameters, where the star ``{\color{blue}$\bigstar$}'' refers to the best fitting point. The scans were performed about the best fit point discussed in this section and the two local minima discussed in Appendix~\ref{sec:app-local-minima}.}
\end{figure}

\begin{figure}[t!]
\centering
\includegraphics[width=1.0\linewidth]{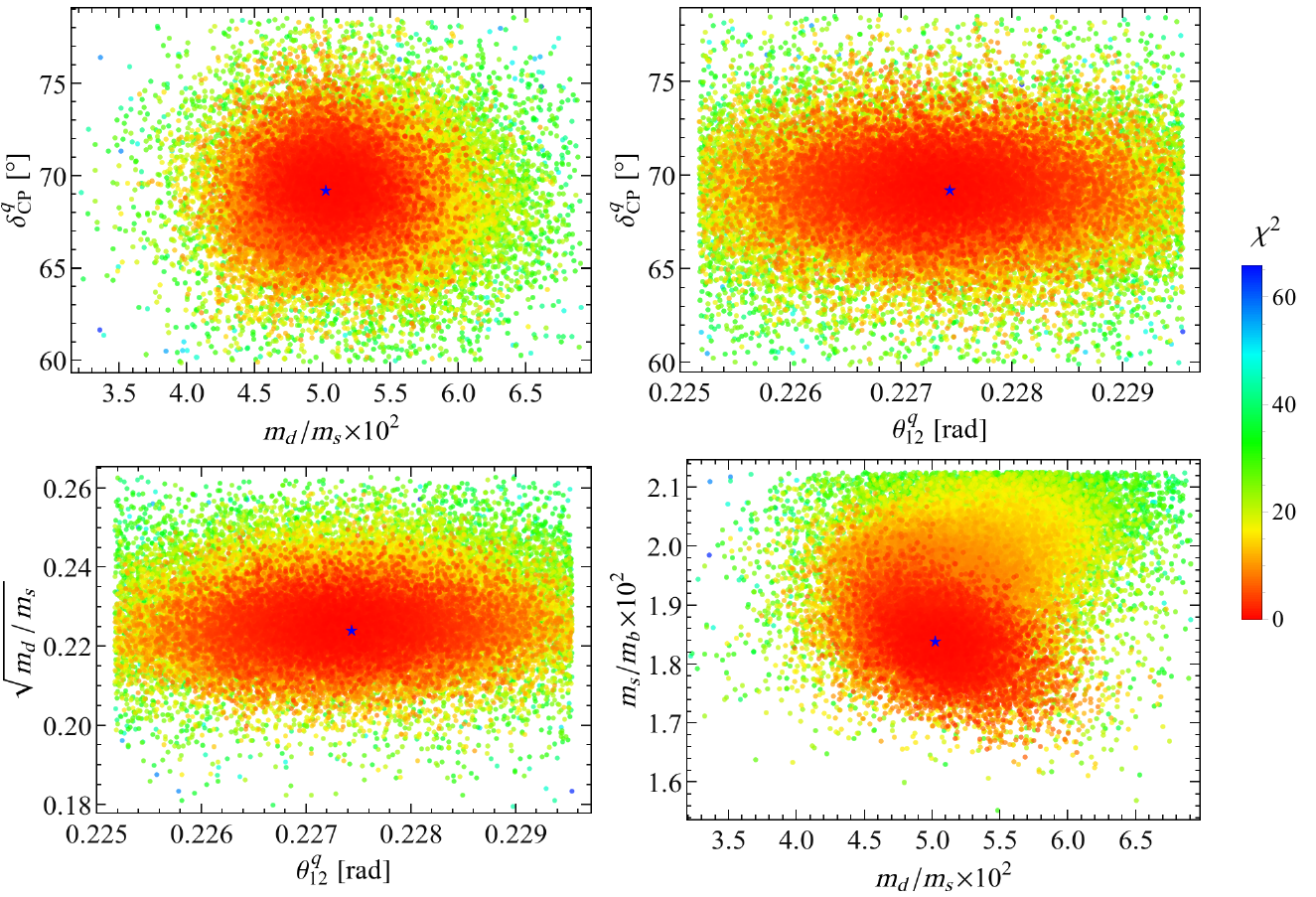}
\caption{\label{fig:correlations-quark} Numerical results for the correlations among the quark mass ratios and mixing parameters, where the star ``{\color{blue}$\bigstar$}'' refers to the best fitting point. The scans were performed about the best fit point discussed in this section and the two local minima discussed in Appendix~\ref{sec:app-local-minima}.}
\end{figure}

We show the contour plots of lepton mixing angles $\sin^2\theta_{12}$, $\sin^2\theta_{13}$, $\sin^2\theta_{23}$ and the mass ratio $m^2_2/m^2_3$ in the plane $r\equiv m_s/m_a$ versus $\eta$ in figure~\ref{fig:contour}, where all other input parameters except $r$ and $\eta$ are fixed to their best fit values collected in table~\ref{tab:best-fit-input}. In the original CSD(3.45) model, there are two small regions in the $\eta-r$ plane which can accommodate the measured values of lepton mixing angles but give CP violation phases of opposite sign. After including the charged lepton corrections, we find only a single small region is phenomenologically viable. Furthermore, we numerically scan the parameter space and require all the obervables in the experimentally preferred $3\sigma$ regions, The correlations between the masses and mixing parameters are shown in figure~\ref{fig:correlations-lepton} and figure~\ref{fig:correlations-quark} for the lepton and quark sectors respectively.

\section{Conclusion}
\label{conclusion}

Modular symmetry offers the possibility to provide an origin of discrete flavour symmetry and to break it along particular symmetry preserving directions, called stabilisers, without introducing flavons or driving fields. The use of multiple moduli at fixed points is justified in the framework of string theory. It is also possible to use weighton fields to account for charged fermion mass hierarchies rather than a Froggatt-Nielsen mechanism. Such an approach can be applied to flavoured GUTs which can be greatly simplified using modular forms.

As an example, we have considered a new modular model based on $(\Gamma_4\simeq S_4)\times SU(5)$, where all flavons and driving fields are removed, with their effect replaced by modular forms with moduli assumed to be at various fixed points, rendering the theory much simpler. The neutrino sector constitutes a minimal 2RHN seesaw model based on CSD($n$) with $n=1+\sqrt{6}\approx 3.45$, intermediate between CSD($3$) and CSD($4$), however being subject to charged lepton corrections. Using the stabilisers, we have reproduced some of the classic features of GUT models such as the GST and GJ relations, although we have seen that these relations apply in a more generalised form as the limiting cases of a choice of parameters.  However, in the case of GJ, this is to be welcomed, since those relations do not work if strictly imposed.

In the considered model we have included a single weighton field to ameliorate the large hierarchies in the charged fermion mass matrices, although some tuning will remain at the per cent level. The best fit to the parameters of the model indicates that the largest charged lepton corrections to CSD($3.45$) mixing are of order the Cabibbo angle, but occurring in both the (1,2) and (2,3) entries of the charged lepton mixing matrix.  Nevertheless the model leads to robust predictions for lepton mixing parameters, which we have compared to those from the pure CSD($3.45$) model with no charged lepton corrections. We have performed a numerical analysis, showing quark and lepton mass and mixing correlations around the best fit points. Since the lightest neutrino mass is zero, and the phases are predicted, the neutrinoless double beta decay parameter is found to be
$m_{\beta\beta}\simeq3.4~\text{meV}$ which is unobservable in the near future.

In conclusion, we have shown that conventional field theory GUT models with flavour symmetry broken by flavons may be drastically simplified using modular symmetry with several moduli assumed to be at their fixed points. The considered $(\Gamma_4\simeq S_4)\times SU(5)$ model results in a dramatic simplification with the 9 flavons and 13 driving fields of the HKL model being replaced by a single weighton and its driving field.

\subsection*{Acknowledgements}

We thank Ferruccio Feruglio for useful comments and Jun-Nan Lu for the help on figure plotting. GJD is supported by the National Natural Science Foundation of China under Grant Nos 11975224, 11835013 and 11947301 and the Key Research Program of the Chinese Academy of Sciences under Grant NO. XDPB15.  S.\,F.\,K. acknowledges the STFC Consolidated Grant ST/L000296/1 and the European Union's Horizon 2020 Research and Innovation programme under Marie Sk\l{}odowska-Curie grant agreement HIDDeN European ITN project (H2020-MSCA-ITN-2019//860881-HIDDeN).  CYY is supported in part by the Grants No.~NSFC-11975130, No.~NSFC-12035008, No.~NSFC-12047533, by the National Key Research and Development Program of China under Grant No. 2017YFA0402200 and the China Post-doctoral Science Foundation under Grant No. 2018M641621.

\section*{Appendix}

\setcounter{equation}{0}
\renewcommand{\theequation}{\thesection.\arabic{equation}}

\begin{appendix}
\section{\label{sec:S4_group_app}Group theory of $\Gamma_4\cong S_{4}$}

The finite modular group $\Gamma_4\equiv\overline{\Gamma}/\overline{\Gamma}(4)$ at level $N=4$ is isomorphic to $S_4$ which is the permutation group $S_4$ of four objects. Geometrically $\Gamma_4$ is the orientation-preserving symmetry group of the cube (or equivalently the octahedron). The group $\Gamma_4$ can be obtained from the generators $S$ and $T$ of the modular group by imposing the additional constraint $T^4=1$, thus they obey the following multiplication rules
\begin{equation}
S^2=(ST)^3=(TS)^3=T^4=1\,.
\end{equation}
The group $\Gamma_4$ has two one-dimensional irreducible representations denoted by $\mathbf{1}$ and $\mathbf{1'}$, a two-dimensional representation denoted by $\mathbf{2}$, and two three-dimensional representations denoted by $\mathbf{3}$ and $\mathbf{3'}$. We adopted the same convention as that of~\cite{Gui-JunDing:2019wap}. We have $S=T=1$ in the trivial singlet representations $\mathbf{1}$ and $S=T=-1$ in the singlet representation $\mathbf{1'}$. For the doublet representation $\mathbf{2}$, the generators $S$ and $T$ are represented by
\begin{equation}
S=\left( \begin{array}{cc}
 0&~1 \\
 1&~0
 \end{array} \right),~~~T=\left( \begin{array}{cc}
 0 ~&~ \omega^2 \\
 \omega ~&~ 0
\end{array} \right)\,,
\end{equation}
with $\omega=e^{2\pi i/3}$. The triplet representations $\mathbf{3}$ and $\mathbf{3'}$ are given by
\begin{equation}
\mathbf{3},~\mathbf{3}':~ S=\pm\frac{1}{3} \left(\begin{array}{ccc}
 1 ~& -2  &~ -2  \\
 -2  ~& -2  &~ 1 \\
 -2  ~& 1 &~ -2
\end{array}\right),~~~T=\pm\frac{1}{3}\left( \begin{array}{ccc}
1 ~& -2\omega^2 &~ -2\omega \\
-2 ~& -2\omega^2 &~ \omega \\
-2  ~& \omega^2 &~ -2\omega
\end{array}\right)\,,
\end{equation}
where the ``+'' and ``$-$'' signs are for $\mathbf{3}$ and $\mathbf{3'}$ respectively. Notice that the triplet representations $\mathbf{3}$ and $\mathbf{3}^{\prime}$ correspond to $\mathbf{3}^{\prime}$ and $\mathbf{3}$ of Refs.~\cite{Penedo:2018nmg,Novichkov:2018ovf} respectively.
In this basis, the tensor products of two doublets $a=(a_1, a_2)^T$ and $b=(b_1, b_2)^T$ are
\begin{equation}
\begin{pmatrix}
a_1 \\
a_2
\end{pmatrix}_{\mathbf{2}}\otimes \begin{pmatrix}
b_1 \\
b_2
\end{pmatrix}_{\mathbf{2}}=\left(a_1b_2+a_2b_1\right)_{\mathbf{1}}\oplus \left(a_1b_2-a_2b_1\right)_{\mathbf{1'}}\oplus \left(\begin{array}{c} a_2 b_2  \\  a_1 b_1 \end{array}\right)_{\mathbf{2}}\,.
\end{equation}
The multiplication laws for the contraction of doublet and triplet are
\begin{eqnarray}
\nonumber\begin{pmatrix}
a_1 \\
a_2
\end{pmatrix}_{\mathbf{2}}\otimes \begin{pmatrix}
b_1 \\
b_2\\
b_3
\end{pmatrix}_{\mathbf{3}}&=&\begin{pmatrix}
a_1b_2+a_2b_3 \\
a_1b_3+a_2b_1  \\
a_1b_1+a_2b_2
\end{pmatrix}_{\mathbf{3}}\oplus \begin{pmatrix}
a_1b_2-a_2b_3 \\
a_1b_3-a_2b_1  \\
a_1b_1-a_2b_2
\end{pmatrix}_{\mathbf{3'}} \,,\\
\begin{pmatrix}
a_1 \\
a_2
\end{pmatrix}_{\mathbf{2}}\otimes \begin{pmatrix}
b_1 \\
b_2\\
b_3
\end{pmatrix}_{\mathbf{3'}}&=& \begin{pmatrix}
a_1b_2-a_2b_3 \\
a_1b_3-a_2b_1  \\
a_1b_1-a_2b_2
\end{pmatrix}_{\mathbf{3}} \oplus \begin{pmatrix}
a_1b_2+a_2b_3 \\
a_1b_3+a_2b_1  \\
a_1b_1+a_2b_2
\end{pmatrix}_{\mathbf{3'}}\,.
\end{eqnarray}
Finally the contraction rules of two triplets are given by
\begin{eqnarray}
\nonumber\begin{pmatrix}
a_1 \\
a_2 \\
a_3
\end{pmatrix}_{\mathbf{3}}\otimes \begin{pmatrix}
b_1 \\
b_2\\
b_3
\end{pmatrix}_{\mathbf{3}}&=&\left(a_1b_1+a_2b_3+a_3 b_2\right)_{\mathbf{1}}\oplus \begin{pmatrix}
a_2 b_2+ a_1 b_3+ a_3 b_1  \\
a_3 b_3+ a_1 b_2+ a_2 b_1
\end{pmatrix}_{\mathbf{2}}  \\
\nonumber&&\oplus \begin{pmatrix}
a_2b_3-a_3b_2 \\
a_1b_2-a_2b_1  \\
a_3b_1-a_1b_3
\end{pmatrix}_{\mathbf{3}}\oplus\begin{pmatrix}
2a_1b_1-a_2b_3-a_3b_2  \\
2a_3b_3-a_1b_2-a_2b_1  \\
2a_2b_2-a_1b_3-a_3b_1
\end{pmatrix}_{\mathbf{3'}} \,,\\
\nonumber\begin{pmatrix}
a_1 \\
a_2 \\
a_3
\end{pmatrix}_{\mathbf{3'}}\otimes \begin{pmatrix}
b_1 \\
b_2\\
b_3
\end{pmatrix}_{\mathbf{3'}}&=&\left(a_1b_1+a_2b_3+a_3 b_2\right)_{\mathbf{1}}\oplus \begin{pmatrix}
a_2 b_2+ a_1 b_3+ a_3 b_1  \\
a_3 b_3+ a_1 b_2+ a_2 b_1
\end{pmatrix}_{\mathbf{2}}  \\
\nonumber&&\oplus \begin{pmatrix}
a_2b_3-a_3b_2 \\
a_1b_2-a_2b_1  \\
a_3b_1-a_1b_3
\end{pmatrix}_{\mathbf{3}}\oplus\begin{pmatrix}
2a_1b_1-a_2b_3-a_3b_2  \\
2a_3b_3-a_1b_2-a_2b_1  \\
2a_2b_2-a_1b_3-a_3b_1
\end{pmatrix}_{\mathbf{3'}} \,,\\
\nonumber\begin{pmatrix}
a_1 \\
a_2 \\
a_3
\end{pmatrix}_{\mathbf{3}}\otimes \begin{pmatrix}
b_1 \\
b_2\\
b_3
\end{pmatrix}_{\mathbf{3'}}&=&\left(a_1b_1+a_2b_3+a_3 b_2\right)_{\mathbf{1}}\oplus \begin{pmatrix}
a_2b_2+a_1b_3+a_3b_1  \\
-(a_3b_3+a_1b_2+a_2b_1)
\end{pmatrix}_{\mathbf{2}}  \\
&&\oplus\begin{pmatrix}
2a_1b_1-a_2b_3-a_3b_2  \\
2a_3b_3-a_1b_2-a_2b_1  \\
2a_2b_2-a_1b_3-a_3b_1
\end{pmatrix}_{\mathbf{3}} \oplus\begin{pmatrix}
a_2b_3-a_3b_2  \\
a_1b_2-a_2b_1  \\
a_3b_1-a_1b_3
\end{pmatrix}_{\mathbf{3'}}\,.
\end{eqnarray}

\section{\label{sec:app-local-minima}Two representative local minima of $\chi^2$}

Since the input parameter space is of high dimension, there are usually some local minima of the $\chi^2$ function besides the global minimum. We give here two  representative local minima with slightly different features, and the best fit values of the input parameters are listed in the following. For the first local minimum, we have
\begin{eqnarray}
\nonumber&&\hskip-0.2in\alpha_{u}/\gamma_{u}=0.05744\,,\beta_{u}/\gamma_{u}=0.4171\,, |\epsilon_{u}/\gamma_{u}|=0.4039\,,\text{arg}(\epsilon_{u}/\gamma_{u})=0.07387\pi\,,\\
\nonumber&&\hskip-0.2in|\alpha_{d1}/\gamma_{d}|=54.6783\,,~\text{arg}(\alpha_{d1}/\gamma_{d})=0.04920\pi\,,
|\alpha_{d2}/\gamma_{d}|=56.3682\,,~\text{arg}(\alpha_{d2}/\gamma_{d})=1.005474\pi\,,\\
\nonumber&&\hskip-0.2in|\beta_{d1}/\gamma_{d}|=2.3851\,,~\text{arg}(\beta_{d1}/\gamma_{d})=0.003877\pi\,,
|\beta_{d2}/\gamma_{d}|=1.9880\,,~\text{arg}(\beta_{d2}/\gamma_{d})=0.9795\pi\,,\\
\nonumber&&\hskip-0.2in|\alpha'_{d1}/\gamma_{d}|=57.8501\,,~\text{arg}(\alpha'_{d1}/\gamma_{d})=1.02773\pi\,,
|\alpha'_{d2}/\gamma_{d}|=58.1969\,,~\text{arg}(\alpha'_{d2}/\gamma_{d})=0.01443\pi\,,\\
\nonumber&&\hskip-0.2in|\beta'_{d1}/\gamma_{d}|=2.2578\,,~\text{arg}(\beta'_{d1}/\gamma_{d})=1.00004372\pi\,,
|\beta'_{d2}/\gamma_{d}|=2.0927\,,~\text{arg}(\beta'_{d2}/\gamma_{d})=1.9788\pi\,,\\
\nonumber&&\hskip-0.2in|\epsilon_{d}/\gamma_{d}|=0.5067\,,~\text{arg}(\epsilon_{d}/\gamma_{d})=1.08775\pi\,,~|\epsilon'_{d}/\gamma_{d}|=0.1507\,,~\text{arg}(\epsilon'_{d}/\gamma_{d})=0.2359\pi\,,\\
&&\hskip-0.2in \gamma'_{d}/\gamma_{d}=0.3441\,,~ m_{s}/m_{a}=0.1357\,,~\eta=1.3321\pi\,.
\end{eqnarray}
For the second local minimum, they are
\begin{eqnarray}
\nonumber&&\hskip-0.2in\alpha_{u}/\gamma_{u}=0.05199\,,\beta_{u}/\gamma_{u}=0.4276\,, |\epsilon_{u}/\gamma_{u}|=0.3923\,,\text{arg}(\epsilon_{u}/\gamma_{u})=1.9618\pi\,,\\
  \nonumber&&\hskip-0.2in|\alpha_{d1}/\gamma_{d}|=54.8850\,,~\text{arg}(\alpha_{d1}/\gamma_{d})=0.05932\pi\,,
|\alpha_{d2}/\gamma_{d}|=60.1485\,,~\text{arg}(\alpha_{d2}/\gamma_{d})=1.0321\pi\,,\\
  \nonumber&&\hskip-0.2in|\beta_{d1}/\gamma_{d}|=2.1891\,,~\text{arg}(\beta_{d1}/\gamma_{d})=0.004274\pi\,,
|\beta_{d2}/\gamma_{d}|=2.1371\,,~\text{arg}(\beta_{d2}/\gamma_{d})=0.9949\pi\,,\\
  \nonumber&&\hskip-0.2in|\alpha'_{d1}/\gamma_{d}|=55.9912\,,~\text{arg}(\alpha'_{d1}/\gamma_{d})=1.03818\pi\,,
|\alpha'_{d2}/\gamma_{d}|=56.1061\,,~\text{arg}(\alpha'_{d2}/\gamma_{d})=0.02192\pi\,,\\
  \nonumber&&\hskip-0.2in|\beta'_{d1}/\gamma_{d}|=2.2935\,,~\text{arg}(\beta'_{d1}/\gamma_{d})=1.01113\pi\,,
|\beta'_{d2}/\gamma_{d}|=2.04182\,,~\text{arg}(\beta'_{d2}/\gamma_{d})=1.9794\pi\,,\\
 \nonumber&&\hskip-0.2in|\epsilon_{d}/\gamma_{d}|=0.5201\,,~\text{arg}(\epsilon_{d}/\gamma_{d})=1.01137\pi\,,~|\epsilon'_{d}/\gamma_{d}|=0.1631\,,~\text{arg}(\epsilon'_{d}/\gamma_{d})=0.04043\pi\,,\\
&&\hskip-0.2in \gamma'_{d}/\gamma_{d}=0.3510\,,~ m_{s}/m_{a}=0.1405\,,~\eta=1.3325\pi\,.
\end{eqnarray}
From the fitted parameters, we can obtain the predictions for the masses and mixing parameters of both quarks and leptons, and the results are summarized in table~\ref{Tab:fitting-local-minima}.

\begin{table}[hpb!]
\centering
\resizebox{1.0\textwidth}{!}{
\begin{tabular}{|c|c|c|c|}
\hline
\hline
\multirow{2}{*}{parameters}
& \multirow{2}{*}{best-fit $\pm$ $1\sigma$} & \multicolumn{2}{c|}{predictions}\\ \cline{3-4}
   & & 1st local minimum & 2nd local minimum\\ \hline
$\theta_{12}^{q}$[rad]
& $0.22736\pm 0.00073$
& 0.22759  & 0.22729 \\
\hline
$\theta_{13}^{q}$[rad]
& $0.00349\pm 0.00013$
& 0.00343 &  0.00349  \\
\hline
$\theta_{23}^{q}$[rad]
& $0.04015\pm 0.00064$
& 0.04050  &  0.03928 \\
\hline
$\delta^{q}_{CP}[^\circ]$
& $69.21330\pm 3.11460$
&  68.95980 & 69.42020 \\
\hline
$m_{u}/m_{c}$
& $(1.9286 \pm 0.6017) \times 10^{-3}$
& $2.0487 \times 10^{-3}$ & $1.8561\times 10^{-3}$ \\
\hline
$m_{c}/m_{t}$
& $(2.8213 \pm 0.1195) \times 10^{-3}$
& $2.7993 \times 10^{-3}$ & $2.7968\times 10^{-3}$ \\
\hline
$m_{d}/m_{s}$
& $(5.0523 \pm 0.6191) \times 10^{-2}$
& $5.0763 \times 10^{-2}$ & $4.9429\times 10^{-2}$ \\
\hline
$m_{s}/m_{b}$
& $(1.8241 \pm 0.1005) \times 10^{-2}$
& $1.8188 \times 10^{-2}$ & $ 1.9564\times 10^{-2}$ \\
  \hline
$m_{t}~[\text{GeV}]$ & $87.45553 \pm 2.08874$ & $ 87.45553$ & $87.45553$ \\
\hline
$m_{b}~[\text{GeV}]$ & $0.96819 \pm 0.01063$ & $0.97177$ & $ 0.98276$ \\
\hline
\hline
$\sin^2\theta_{12}^{l}$ (NO)
&$0.304_{-0.012}^{+0.012}$ & $0.30721$ & $0.31340$ \\
\hline
$\sin^2\theta_{13}^{l}$ (NO)
& $0.02219_{-0.00063}^{+0.00062}$ & $0.02231$ & $0.02242$ \\
\hline
$\sin^2\theta_{23}^{l}$ (NO)
& $0.573_{-0.020}^{+0.016}$ & $0.58286$ & $0.57132$ \\
\hline
$\delta_{CP}^{l}[^\circ]$ (NO)
& $197_{-24}^{+27}$ & $158.22542$ & $154.42225$ \\
\hline
$m_{e}/m_{\mu}$ & $(4.73689 \pm 0.04019) \times 10^{-3}$ & $4.73926 \times 10^{-3}$  & $ 4.73272\times 10^{-3}$ \\
\hline
$m_{\mu}/m_{\tau}$ & $(5.85684 \pm 0.04654) \times 10^{-2}$ & $5.84221 \times 10^{-2}$  & $ 5.87193\times 10^{-2}$\\
\hline
$m_{\tau}~[\text{GeV}]$ & $1.30234 \pm 0.00679$ & $1.30234$ & $1.30234$\\
  \hline
$\Delta m_{21}^{2}~[10^{-5}\text{eV}^{2}]$ (NO)
&  $7.42_{-0.20}^{+0.21}$ & $7.42000$ & $7.42000$ \\
\hline
$|\Delta m_{31}^{2}|~[10^{-3}\text{eV}^{2}]$ (NO)
&  $2.517_{-0.028}^{+0.026}$ & $2.49889$ & $2.51657$ \\
\hline
$\rho/^{\circ}$ (NO)
& $-$ & $9.21346$ & $7.13008$ \\
\hline
$m_1 [\mathrm{meV}]$ (NO)
& $-$ & $0$ & $0$ \\
\hline
$m_2 [\mathrm{meV}]$ (NO)
& $-$ & $8.61394$ & $8.61394$ \\
\hline
$m_3 [\mathrm{meV}]$ (NO)
& $-$ & $49.98889$ & $50.16546$ \\ \hline

$\chi^2_{\text{min}}$  &  &  $4.1205$  & $9.2244$ \\
\hline
\hline
\end{tabular}}
\caption{\label{Tab:fitting-local-minima} The predicted values of the masses and mixing parameters of quark and lepton at the two representative local minima of $\chi^2$.
}
\end{table}

\end{appendix}


\providecommand{\href}[2]{#2}\begingroup\raggedright\endgroup

\end{document}